\documentclass{aa}
\usepackage{graphics}
\usepackage{natbib}
\usepackage{times}\fontfamily{ptm}\selectfont

\def\etal{et al.\/}
\def\cf{{\em cf.\/}}
\def\ie{{\em i.e.\/}}

\def\Msun{\mbox{$M_{\sun}$}}

\begin{document}
\hbadness 10000

   \thesaurus{05(10.07.2; 10.11.1; 10.07.03 M 55; 08.12.3)} 

   \title{The Stellar Distribution of the Globular Cluster M55
          \thanks{Based on observations made at the European Southern
                  Observatory, La~Silla, Chile.} }

   \author{Simone R. Zaggia   \inst{1,2} \and 
           Giampaolo Piotto   \inst{1}   \and 
           Massimo Capaccioli \inst{2,3} }

   \offprints{S.R. Zaggia.
              e-mail: zaggia@na.astro.it}

   \institute{  Dipartimento di Astronomia, Universit\`a di Padova,
                Vicolo dell' Osservatorio 5, I--35122 Padova, Italy.
           \and Osservatorio Astronomico di Capodimonte, via Moiariello 16,
                I-80131 Napoli, Italy.
           \and Dipartimento di Scienze Fisiche, Universit\`a Federico~II, 
                Mostra d'Oltremare, Padiglione 19, I-80125 Napoli, Italy.}

   \date{Received 29 May 1997 / Accepted 11 July 1997}
%   \date{Received  / Accepted }

   \maketitle

   \begin{abstract} We have used extensive $V$, $I$ photometry (do\-wn to
$V=20.9$) of $33615$ stars in the direction of the globular cluster
M55 to study the dynamical interaction of this cluster with the tidal
fields of the Galaxy. An entire quadrant of the cluster has been covered, out 
to $\simeq1.5$ times the tidal radius.

A CMD down to about 4 magnitudes below the turn-off is presented and
analysed. A large population of BS has been identified. The BS are
significantly more concentrated than the other cluster stars in the inner 300
arcsec, while they become less concentrated in the cluster envelope.

We have obtained luminosity functions at various radial intervals from
the center and their corresponding mass functions.  Both clearly show
the presence of mass segregation inside the cluster. A dynamical
analysis shows that the observed mass segregation is compatible with
what is predicted by multi-mass Ki\-ng-Michie models.  The global mass
function is very flat with a po\-wer-law slope of $x=-1.0\pm0.4$.  This
suggest that M55 might have suffered selective losses of stars, caused
by tidal interactions with the Galactic disk and bulge.

The radial density profile of M55 out to $\sim 2\times r_t$ suggests the
presence of extra-tidal stars whose nature could be connected with the
cluster.

\end{abstract}

%________________________________________________________________

\section{Introduction}

The recent advances in our understanding of the structure and evolution
of Galactic globular clusters (GCs) have been possible thanks to the
advent of accurate CCD photometry.
However, till few years ago, CCD photometry was limited to the 
internal parts of GCs due to the small fields of the detectors.
All the information relative to the outer regions and to the tidal
radius $r_t$, arise from visual (\emph{by eye}) stellar counts made on
Schmidt plates, especially by King and collaborators
\cite[]{King68,Trager95}.
This methodology of investigation suffers from various problems and 
statistical biases;
we list some of them:
\begin{itemize}
\item The limiting magnitude of photographic plates, which is ge\-nerally
too bright to permit the investigation of the radial distribution of
stars in an appropriate mass range;
\item the high uncertainty in the evaluation of background stellar
contamination;
\item an insufficient crowding/completeness correction.
\end{itemize}

All the more recent models of dynamical evolution need to make 
assumptions on the mass function, on the effects due to the radial 
anisotropy of the velocity distribution, and on the mass segregation 
which, in principle, could be determined observationally.

\begin{figure*}
\resizebox{12cm}{!}{\includegraphics{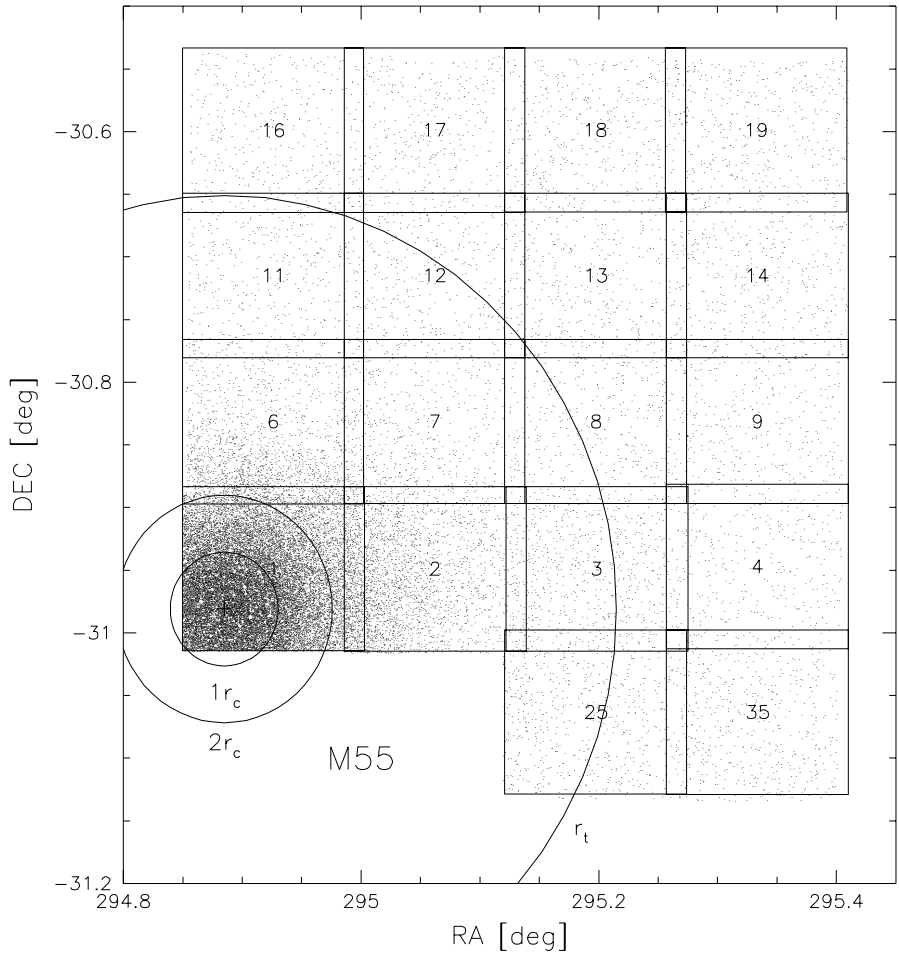}} 
\hfill
\parbox[b]{55mm}{
\caption{EMMI fields coverage of M55. 
The center of the cluster is inside field number 1.
The circles mark the approximate position of: one core radius, $1r_c$, 
two core radii, $2r_c$, and the tidal radius, $r_t$.}}
\label{M55frames}
\end{figure*}

Fostered by this lack of observational data, five years ago we started a
long term project using one of the largest field CCD cameras available,
EMMI at the NTT, to obtain accurate stellar photometry in two bands,
$V$ and $I$, over at least a full quadrant in a number of GCs.
The sample was selected taking into account the different 
morphological types and the different positions in the Galaxy.
The principal aim was to map the stellar distribution from the central
part out to the outer envelope (beyond the formal tidal radius, for a
better estimate of the field stars contamination and in order to
investigate on the possible presence of tidal tails), with a good
statistical sampling of the stars in distinct zones of the color
magnitude diagram (CMD) and of different masses.

The use of CCD star counts, instead of photographic Schmidt plates for which
the only advantage is still to give a larger area coverage 
\cite[for works on this subject]{Grillmair95,LS97}, allows us to go
deeper inside the core of the clusters, to better handle photometric
errors and completeness corrections and to reach a considerably fainter
magnitude level.
CCDs also allows in the case of high concentration clusters 
to complement star counts of the central part with aperture photometry
of short exposures images \cite[]{Ivo97}.

An important byproduct of this study is the photometry of a
significant number of stars in all the principal sectors of
the CMD.
This sample is of fundamental importance to test modern
evolutionary stellar models \cite[]{RFP88}.
In all cases the CMD extends well below the turn off of the main 
sequence.
This permits us to estimate the effect of mass segregation for masses
from the TO mass ($\sim0.8\Msun$) down to $0.6, 0.5\Msun$.

So far, we have collected data for a total of 19 clusters.  
Same of them have already been reduced and analyzed.
In this work we present the analysis of the star counts of the globular 
cluster \object{NGC 6809}=M55.
Other clusters, for which we have already given a first report elsewhere
\cite[]{Zaggia95,Carla96,Ivo95,Alf96}, will be presented in future works 
\cite[]{Ivo97,Alf97}.

\section{Why the globular cluster M55?}
M55 is a low central concentration, $c=0.8$ \cite[]{Trager95}, low
metallicity, ${\rm [Fe/H]}=-1.89$ \cite[]{Zinn80}, cluster located at
$\simeq4.9$~kpc from the Sun \cite[]{Mand96}.
Although it is a nearby object, it has received little or sporadic
attention until very recently.
The works of \cite{Mateo96} and \cite{Fahlman96} presented
photometric datasets of M55 that have been used principally to establish 
the age and the tidal extension of the Sagittarius dwarf 
galaxy.
\cite{Mand96} published the first deep (down to $V\simeq24.5$)
photometry of the cluster (other previous studies of the stellar
population of M55 are in Lee, 1977\nocite{Lee77};
Shade, VandenBerg, and Hartwick, 1988\nocite{Shade88}; 
and Alcaino \etal, 1992\nocite{Alcaino92}).
From the data of a field at $\simeq2$~core radii from the center,
\cite{Mand96} estimated a new apparent distance modulus for
M55, $(m-M)_V=13.90\pm0.09$, and from the luminosity function they found
that the high-mass end of the mass function ($0.5<M/M_{\odot}<0.8$) is
well fitted by a power law with $x=0.5\pm0.2$, whereas at the low-mass
end ($M/M_{\odot}<0.4$) the mass function has a slope of $x=1.6\pm0.1$.

From the dynamical point of view, M55 has been previously studied by
\cite{Pryor91} in their papers on the mass-to-light ratio of
globular clusters.
Their principal conclusion is that this cluster might have a power law 
mass function with an exponent $x=1.35\div2.0$, with a lower limit of
the mass function in the range $\simeq0.1\div0.3\Msun$ (\ie\ a total
absence of low mass stars): a conclusion opposite to that found recently
by \cite{Mand96}.

\begin{figure*}
\resizebox{12cm}{!}{\includegraphics{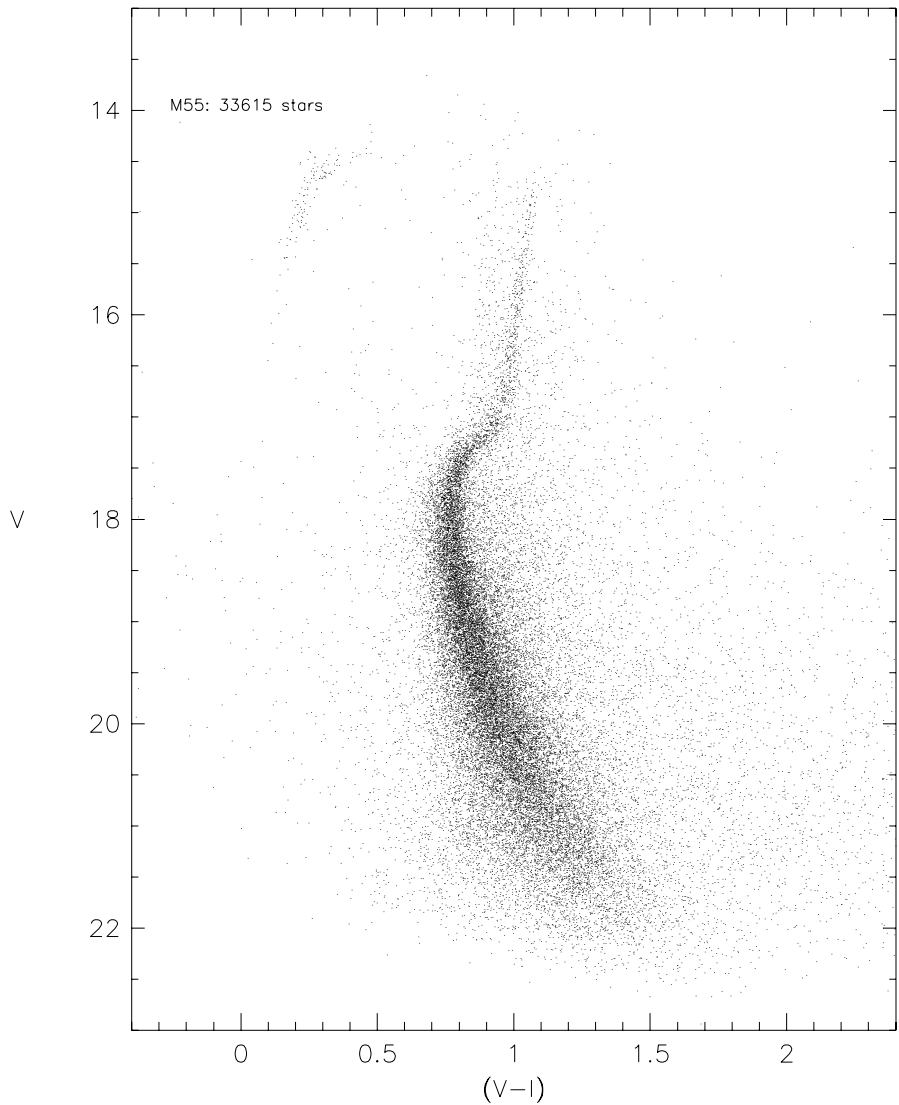}}
\hfill
\parbox[b]{55mm}{
\caption{Color magnitude diagram $V vs. (V-I)$ for 33615 of M55.}}
\label{M55tot}
\end{figure*}

An original work on M55 is in \cite{Irwin84}.
They studied the radial star count density profile using photographic 
material digitalized with the \emph{Automatic Pla\-te
Measuring System} (APM) of the Cambridge University.
\cite{Irwin84} used a single photometric band, which
did not allow them to lower the contribution of the field stars in the
construction of the radial star counts.
Nevertheless in this work (never repeated in other clusters), the
authors reach some interesting conclusions: they claim one of the first
evidences of mass segregation (even if they cannot quantify it); the
central stellar luminosity function seems to be flat (with a
corresponding mass function having a slope of $x\simeq0.0$) with a
partial deviation from the King models.
Moreover, they claim the presence of 8 short period variables, at the limit 
of their photometry, compatible with contact binaries of W~UMa type.
This last point is interesting for the presence of a large population of 
Blue Straggler (BS) stars in M55 to which the variables of 
\cite{Irwin84} could belong.

Despite the potential interest of this ne\-arby cluster
for problems such as the dynamical evolution of globular clusters and
interaction with the tidal field of the Galaxy, the existing data
on M55 are so far limited and have been used to address only
particular problems.  Now large field CCDs offer the possibility to
attack this problems in a suitable way.  
The following Section is
dedicated to the presentation of the M55 data set and our observing
strategy; in Section~4 we show the luminosity and mass function of the
cluster; in Section~5 we present the analysis of the radial density
profile and the conclusions.  The details of the techniques adopted in
the reduction and analysis of the data can be found in the appendix of
the paper.

\begin{figure*}
\resizebox{\hsize}{!}{\includegraphics{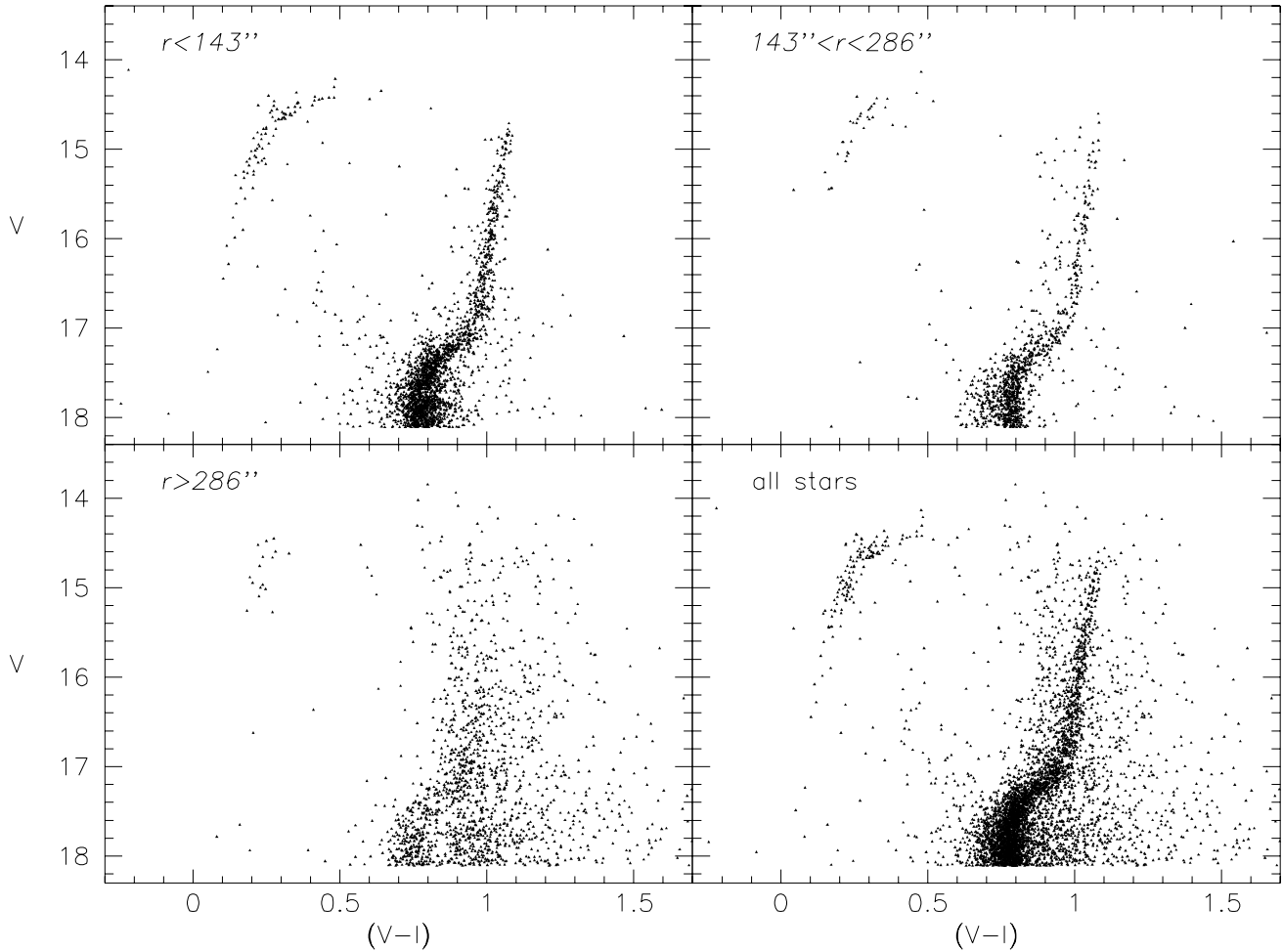}}
\caption{CMD $V vs. (V-I)$ of M55 stars in three different radial
intervals corresponding to $r\le r_c$, $r_c <r \le 2r_c$, and $r>2r_c$.}
\label{cmdradial}
\end{figure*}

\section{The photometric data}
A whole quadrant of M55 was mapped (from the center out to
$\sim1.5~r_t$, with $r_t=977''$ as in Trager, King, and Djorgovski,
1995\nocite{Trager95}), on the night of July~5~1992 with 18~EMMI-NTT fields
($7\farcm2\times7\farcm2$) in the $V$ and $I$ bands.
Figure~\ref{M55frames} shows the field positions on the
sky. For each field a $V$ and a $I$ band image were taken in succession,
with exposure times of 40 and 30 seconds respectively.  The night was
not photometric and the observing conditions improved as we moved from the
outer fields to the internal ones. Information on the
various fields and on all the technicalities of the reduction and
analysis are reported in the appendix of the paper.

The $V$ \emph{vs.}\/ $(V-I)$ color magnitude diagram for a total of 33615 
stars of M55 is shown in Figures~\ref{M55tot} and \ref{cmdradial}.
In total we detected 36800 objects of the cluster$+$field; 
$\simeq9\%$ of them were eliminated after having applied a selection in the 
{\tt DAOPHOT II\/} PSF interpolation parameters as in \cite{Piotto90a}.
Although the exposure time was relatively short, 
the brightest stars of the red giant branch and of the 
asymptotic giant branch are saturated, though they can be still used for the 
radial star counts.
We have omitted them from the final CMD.

In the following we will analyze the data using a division 
into three radial subsamples: inner ($r\le r_c$),
intermediate ($r_c<r\le 2r_c$), and outer ($2r_c<r\le r_t$). 
The core radius is $r_c=143\arcsec$, as found from the radial density 
profile analysis (\cf\ Section~5).
In Figure~\ref{cmdradial} we show the brightest part of the CMD of M55,
divided in the three radial subsamples.
A large population of blue straggler stars (BS) is clearly visible,
particularly in the inner part of the cluster where the
background/foreground star contamination is low.
In the intermediate zone, the BS population is better defined, and the 
sequence seems to reach brighter magnitudes.
The BS sequence of the inner part appears to be broader in color than 
the sequence of the intermediate radial range.
Part of this broadening can be attributed to the photometric errors that 
are larger in the inner region than in the intermediate one.
The rest of the broadening is probably natural and could be connected to
the two formation mechanisms of BS stars: the outer BS stars might
mainly come from merger events, while the inner BS might be the final
products of collisions (see Bailyn 1995\nocite{Bailyn95}). 

\begin{figure}
\resizebox{\hsize}{!}{\includegraphics{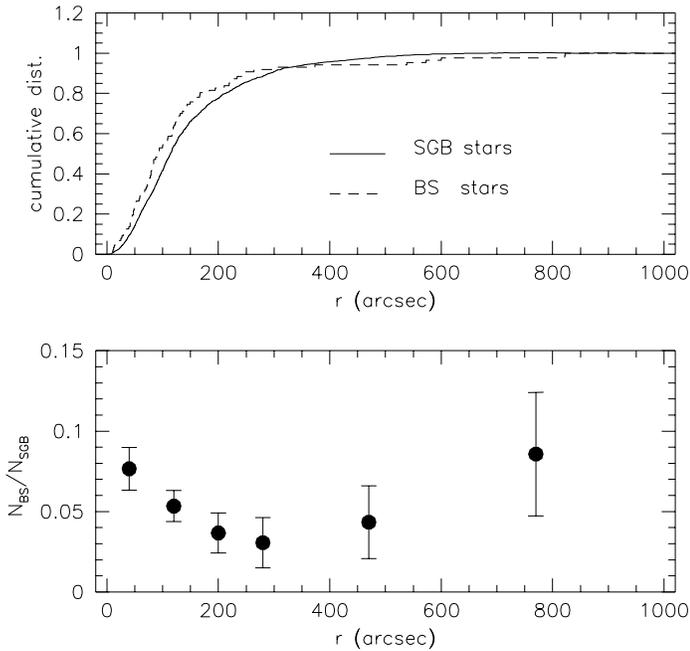}}
\caption{\emph{Upper panel.} Cumulative distribution of the SGB stars 
(solid line) and of BS stars (dashed line).  \emph{Lower panel.} Radial
trend of the ratio of the number of BS stars ($N_{BS}$) and the number of 
SGB stars ($N_{SGB}$) in the same magnitude interval as the BS. Note the
bimodal distribution.}
\label{figbs}
\end{figure}

We compared the distribution of the 95 BS with that of the 
1669 sub-giant branch (SGB) stars selected in the same magnitude interval.
In order to minimize the background star contamination along the SGB
(very low indeed in the inner part of the cluster), we chose only the
stars inside $\pm2.0\sigma$ (where $\sigma$ is the standard deviation in
the mean color) from the mean position of the SGB.
We subtracted the background stellar contamination estimated from the
star counts in the radial zone $r>1.1~r_t$.
The BS seem to be more concentrated than the corresponding SGB stars
only in the inner $250\div300\arcsec$ (Figure~\ref{figbs}). At larger
distances, the BS distribution becomes less
concentrated than the comparison SGB stars. We run a 2-population
Kolmogorov-Smirnov test. The test does not give a particularly high
statistical significance to the result: the probability that the BS
and SGB stars are \emph{not} taken from the the same distribution is 
96\%.  However, this possibility cannot be excluded: see 
\cite{Piotto90b} and \cite{DP93} for a discussion on the
limits in applying this statistical test for checking population
gradients. Another way to look into the same problem is to investigate the
radial trend of the ratio BS/SGB, as plotted in Figure~\ref{figbs} (\emph{lower
panel}). Also in this case the bimodal trend is quite evident. The relative
number of BS stars decreases from the center of the cluster to reach a
minimum at $r\simeq 250\div300$ arcsec ($r\simeq 2r_c$), and then it rises again.
Again, the statistical significance is questionable, in view of the small 
number of BS at $r>300$ arcsec (6 stars). Nevertheless, this possible
bimodality is noteworthy.
Indeed, there is a growing body of evidence that the radial
distribution of BS stars in GCs might be bimodal, as shown by 
\cite{Ferraro97} for \object{M3} or \cite{Ivo97} and
\cite{Piotto97} for \object{NGC 1851}. What makes our result for M55 
of some interest is that this distribution has been interpreted in terms
of environmental effects on the production of BS stars. However, the fact
that M55 has a very low concentration (c=0.8), while M3 and NGC 1851
are high concentration clusters (c=1.85 and c=2.24 respectively, 
Djorgovski 1993\nocite{DJ93}), might make this conclusion at least
questionable.

\section{Luminosity and mass function \label{M55lf}}
\begin{figure*}
\resizebox{12cm}{!}{\includegraphics{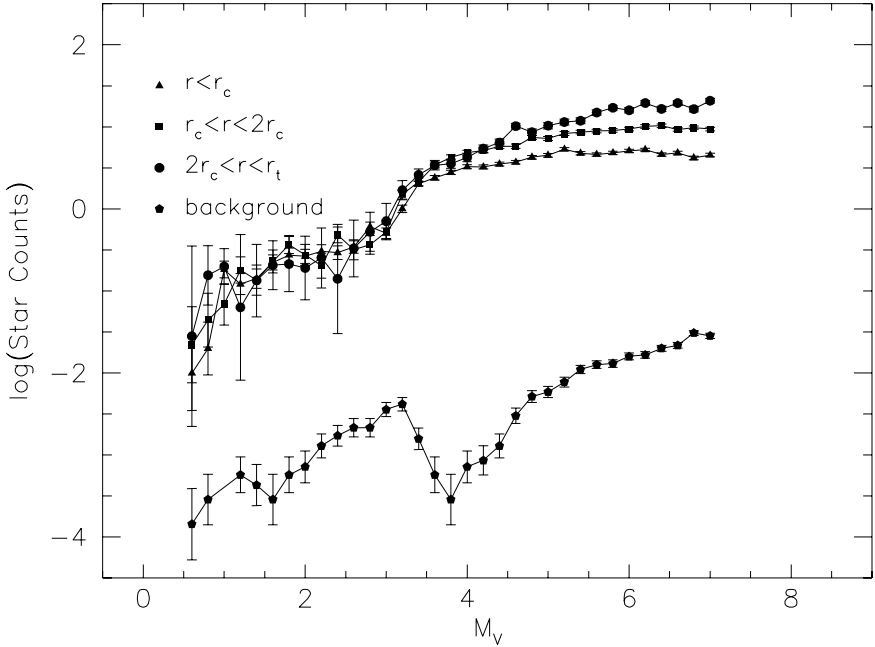}}
\hfill
\parbox[b]{55mm}{
\caption{Stellar luminosity functions for the inner, $r\le r_c$,
intermediate, $r_c<r\le 2r_c$, and outer, $2r_c<r\le r_t$, annuli of M55.
The color-selected field luminosity function (filled hexagons) is
vertically shifted down for clarity.}}
\label{M55fdl}
\end{figure*}

From the CMD we have derived a luminosity function (LF) for the stars of 
M55.
Figure~\ref{M55fdl} shows the LFs in the different annuli defined in
the previous Section (inner, intermediate and outer). 

The three LFs have been normalized to the star counts of the SGB region 
in the magnitude interval $15.90<V<17.40$,
after subtracting the contribution of the background/foreground stars
scaled to the area of each annulus.
In the lower part of Figure~\ref{M55fdl}, we show also the LF of the 
background/foreground stars estimated from the star counts at
$r>1.3~r_t$ vertically shifted for clarity.
In order to reduce contamination by those stars, all the LFs have been
calculated selecting the stars within  $2.5\sigma$ (again, $\sigma$ is
the standard deviation of the mean color) from the fiducial line of the
main sequence of the cluster.
The LFs do not include the HB and BS stars.
The LF of the background stars has a particular shape: it suddenly drops
at $M_V=4.0$.
This feature has a natural explanation considering the color-magnitude 
distribution of the field stars around M55 and the way we selected the stars.
The drop in the number of field stars is at the level of the M55 TO and as 
can be seen in Figure~\ref{M55tot}, or in 
the lower right panel in Figure~\ref{cmdradial}, the TO of M55 is bluer than
the TO of the halo stars of the Galaxy, which are the main components of
the field stars towards M55 \cite[]{Mand96}.
Selecting only stars within $2.5\sigma$ of the fiducial line of M55 
will naturally cause such a drop.

The completeness correction, as obtained in ap\-pen\-dix~\ref{crowd}, has been
applied to the stellar counts of each field of M55.
As it is possible to see from Table~\ref{M55tab2}, the magnitude limit 
varies from field to field.
We have adopted the same, global, magnitude limit for all the LFs: \ie, 
that of the fields with the brighter completeness limit (field 16 and 17).
This limits all the LFs to $V=20.9$, corresponding to a stellar mass 
$m\sim0.6\Msun$, for the adopted distance modulus and a standard 15~Gyr 
isochrone (see next subsection).
The data for the inner annuli come from the central image, which has a
limiting magnitude of the corresponding LF fainter than the global value
adopted here.
This is due to the better seeing of the central image 
compared to all the other images.
We adopted a brighter limiting magnitude in order to avoid problems in comparing the
different LFs.

Figure~\ref{M55fdl} shows clearly different behaviour of the LFs below the 
TO: they are similar for the stars above the TO, while
the LFs become steeper and steeper from the inner to the
outer part of the cluster: this is a clear sign of mass segregation.
For the inner LF there is also a possible reversal in slope 
below $M_V=5.5$.

In order to verify that the difference between the three LFs is not due to 
systematic errors (wrong completeness correction, imperfect combination of
data coming from two adjacent fields etc.), we have tested our combining
procedure in several ways.
In one of our tests we built LFs of two EMMI fields at the same distance
from the center of the cluster: \ie, we compared the LF of the field~2 
with that of the field~6.
After having corrected for the ratio between the covered areas and
subtracting the field star contribution, the two LFs were
consistent in all the magnitude intervals down to the completeness
level of the data (that is lower than the one adopted).
Having for field~2 a magnitude limit of 22.2 (see
Table~\ref{M55tab2}) and field~6 a limit of 21.5, we also verified that
for the latter our star counts are in correct proportion below the
completeness level of 50\%.

In a second test, we generated two LFs dividing the whole cluster in two
octants (dividing along the $45^\circ$ line that runs from the center of
the cluster till the field 19 \cf\ Figure~\ref{M55frames}).
For each of the two slices we generated three LFs in the same radial
range as in Figure~\ref{M55fdl}.
After comparing all of them we did not find any significant difference.
Therefore the differences among the three LFs in Figure~\ref{M55fdl}
must be real.

\begin{figure*}[t]
\resizebox{12cm}{!}{\includegraphics{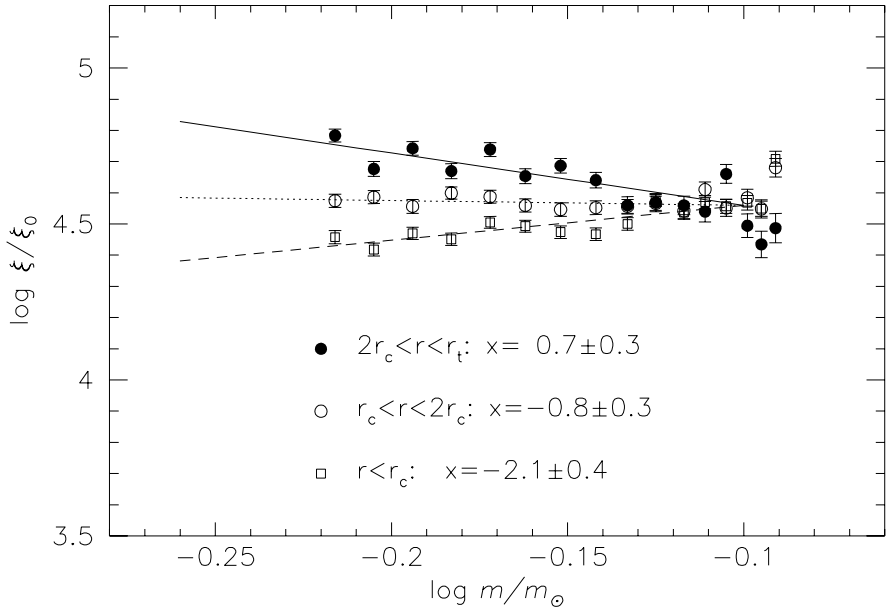}}
\hfill
\parbox[b]{55mm}{
\caption{Mass functions for M55 for three radial ranges: 
inner, $r\le r_c$, intermediate, $r_c<r\le 2r_c$, and outer, 
$2r_c<r\le r_t$.
The effect of mass segregation is clearly visible.
The slope $x$ corresponds to the index $x$ of the power law
$\xi=\xi_0m^{-(1+x)}$ fitted to the data in the range
$-0.23<\log(m/m_\odot)<-0.1$.}}
\label{M55mf}
\end{figure*}

Another source of error in the LF construction is represented by the 
LF of the field stars. 
As will be shown in Section~\ref{M55rprof}, M55 has a halo of probably 
unbound cluster stars.
The field star LF constructed from the star counts just outside 
the cluster can be affected by some contamination of the cluster halo.
The consequence is that we might over-subtract stars when subtracting the
field LF from the cluster LF, modifying in this way the slope of the
mass function (the more affected magnitudes are the faintest ones).
To test this possibility, we have extracted background LFs in two
different anulii outside the cluster (in terms of $r_t$, $1.0<r\le1.3$ and 
$r>1.3$).
Comparing the two background/foregr\-ou\-nd LFs we found that the number of
stars probably belonging to the cluster but outside the tidal radius
must be less than $\sim25\%$ of the adopted field stars in the worst
case (the faintest bins).
The possible over-subtraction is not a problem for the inner and
intermediate LFs, where the number of field stars (after rescaling for
the covered area) is always less than $\sim3\%$ of the stars counted in
each magnitude bin.
For the outer LF, the total contribution of the measured field stars
is larger, but it is still less than $25\%$ of the cluster stars (the
worst case applies to the faintest magnitude bin): this means that the
possible M55 halo star over-subtraction in the field-corrected LF is always
less than $6\%$ ($25\% \times 25\%$), negligible for our purposes.

\subsection{Mass function of M55 \label{Sm55mf}}
In order to build a mass function for the stars of M55, we needed to adopt a
distance modulus and an extinction coefficient.
\cite{Shade88} give $(m-M)_V=14.10$,
E$(B-V)=0.14\pm0.02$, while, more recently, \cite{Mand96} give
$(m-M)_V=13.90\pm0.07$, E$(B-V)=0.14\pm0.02$.
In the absence of an independent measure made by us, we adopted the 
values published by \cite{Mand96} because they are based on the 
application, with updated data, of the subdwarfs fitting method.
Using the LFs of the previous Section we build the corresponding mass
functions using the mass-luminosity relation tabulated by \cite{VDB85}
for an isochrone of $Z=3\times10^{-4}$ and an age of
16~Gyr \cite{Alcaino92}.
The MFs for the three radial intervals are presented in
Figure~\ref{M55mf}.
The MFs are vertically shifted in order to make their comparison more 
clear.

The MFs are significantly different: the slopes of the MFs increase
moving outwards as expected from the effects of the mass segregation and 
from the LFs of Figure~\ref{M55fdl}.
Figure~\ref{M55mf} clearly shows that the MF starting from the center
out to the outer envelope of the cluster is flat: the index $x$ of the
power law, $\xi=\xi_0m^{-(1+x)}$, best fitting the data are:
$x=-2.1\pm0.4$, $x=-0.8\pm0.3$, and $x=0.7\pm0.4$ going from the
inner to the outer anulii;
this means that the slope of the global MF (of all the stars in M55)
should be extremely flat.
Indeed, the slope of the global mass function obtained from the
corresponding LF of all the stars of M55 is: $x=-1.0\pm0.4$ 
This result agrees with the results of \cite{Irwin84},
while the results of \cite{Pryor91} appear in contrast to what we
have found here.

Our MF in the outer radial bin can be compared with the high-mass MF of
\cite{Mand96}, obtained from a field located at
$\simeq6$~arcmin from the center of M55.
As already reported in Section~2, \cite{Mand96} obtained a
deep MF for M55 (down to $M\simeq0.1\Msun$) which they describe 
with two power laws connected at $M\simeq 0.4\div0.5\Msun$.
Their value of $x=0.5\pm0.2$ for the high-mass end of the mass function
($0.5<M/\Msun<0.8$) is in good agreement with our value of
$x=0.7\pm0.4$, obtained in the same mass range for the outer radial bin. 
The low-mass end of the MF by \cite{Mand96} ($M/\Msun<0.4$) 
has a slope of $x=1.6\pm0.1$.

The level of mass segregation of M55 is comparable to that found in 
\object{M71} by \cite{Richer89}.
M71 shares with M55 similar structural parameters as well as positional
parameters inside the Galaxy.
The detailed analysis of \cite{Richer89} of M71 showed that
this cluster should also have a large population of very low mass stars
($\sim0.1$\Msun).

By fitting a multi-mass isotropic King model \cite[]{King66,GG79}
to the observed star density profile of M55, we compared the observed
mass segregation effects with the one predicted by the models.
Here we give a brief description of our assumptions in order to
calculate the mass segregation correction from multi-mass King models. 
A more detailed description can be found in \cite{Pryor91}, from 
which we have taken the \emph{recipe}.
The main concern in the process of building a multi-mass model is in the 
adoption of a realistic global MF for the cluster.
For M55 we adopted a global MF divided in three parts: 
\begin{itemize}
\item a power-law for the low-mass end, $0.1<M/\Msun\le 0.5$, 
with a fixed slope of $x=1.6$ (as found by Mandushev \etal\ 
1996\nocite{Mand96});
\item a power-law for the high-mass end, $0.5<M/\Msun\le m_{TO}$, with a
variable slope $x$;
\item and a power-law for the mass bins of the dark-remnants where to
put all the evolved stars with mass above the TO mass,
$m_{TO}<M/\Msun\le 8.0$: essentially white dwarfs.
Here we adopted a fixed slope of 1.35,  
The mass of the WDs were set according to the initial-final mass
relation of \cite{W90}.
\end{itemize}
\begin{figure}[t]
\resizebox{\hsize}{!}{\includegraphics{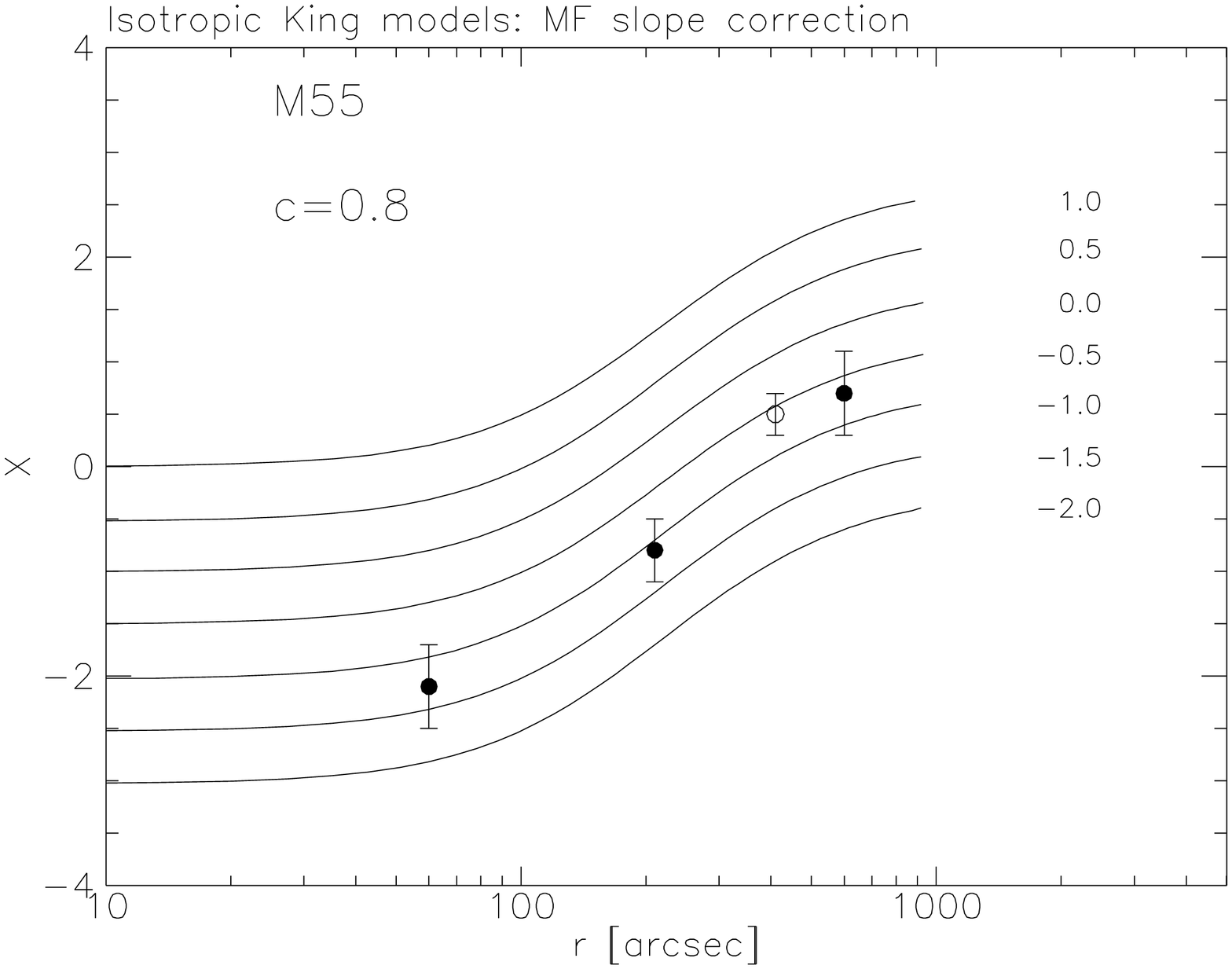}}
\caption{
Isotropic King model MF slope correction for M55. 
The full dots are the slopes of the MFs obtained in this paper, the open
circle is the measure of the high-end MF of Manddushev \etal, 1996.}
\label{M55slx}
\end{figure}

\begin{figure}[t]
\resizebox{\hsize}{!}{\includegraphics{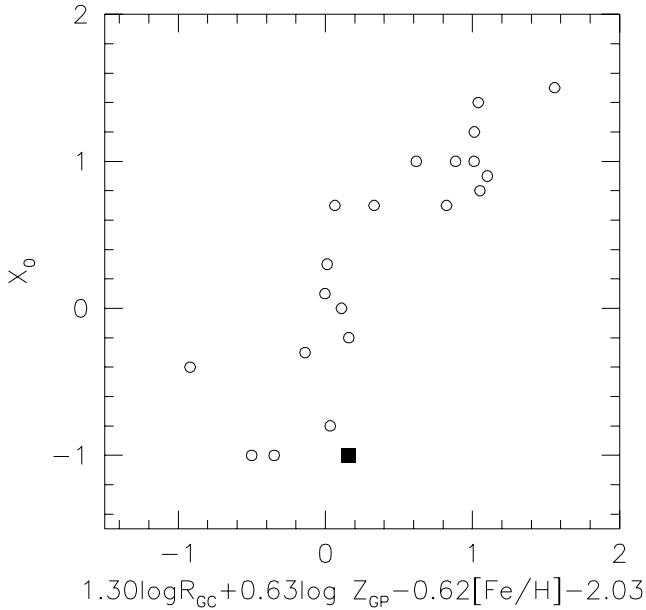}}
\caption{Trivariate relation from Zoccali \etal (1997), between the 
distance from the Galactic center ($R_{GC}$), the height above the 
Galactic plane ($Z_{GP}$), the metallicity of the cluster ([Fe/H]), and
the slope of the global stellar mass function ($X_0$) of a globular
cluster. The filled square marks the position of M55.}
\label{m55xrz}
\end{figure}
\begin{figure}[t]
\resizebox{\hsize}{!}{\includegraphics{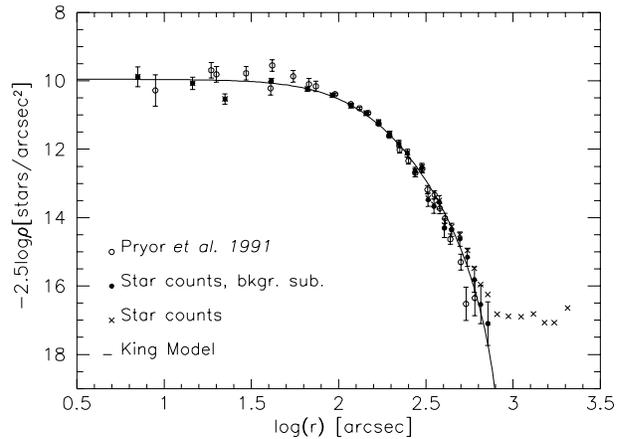}}
\caption{Radial density profile of M55.
Small crosses represent the raw stellar counts;
filled dots are our star counts after subtracting the background 
counts contribution;
open dots shows the M55 profile published by Pryor \etal (1991);
the continuous line is the single mass King model fitted to our star counts
($c=0.83,\, r_c=143''$).}
\label{profile}
\end{figure}

To build the mass segregation curves we varied the MF slope $x$ 
(the only variable parameter of the models)
of the high-mass end stars in the range $-1.0\div1.35$, finding for each
slope the model best fitting the radial density profile of the cluster.
Then we calculated the radial variation of $x$ for the best-fit models 
in the same mass range of the observed stars: $0.5<M/M_{\odot}<m_{TO}$.
The radial variations of $x$ are compared with the observed MFs in
Figure~\ref{M55slx}.
The mass function slopes are shown at the right end of each curve.
This plot is similar to those presented in \cite{Pryor86}, 
and allows one to obtain the value of the global mass function of
the cluster.
The three observed points follow fairly well the theoretical curves.
Also the high-mass MF slope value of \cite{Mand96} (open circle in
Figure~\ref{M55slx}) is in good agreement with the models and our MFs.
From these curves, we have that the slope of the high-mass end of the 
global MF of M55 is $x\simeq-1.0$, which is in quite good agreement with 
the global value of the MF found from the global LF of M55 (\cf\ previous
section).
In Figure~\ref{profile} we show the model which best fits the observed
radial density profile for a global mass function with a slope $x=-1.0$.

The relatively flat MF of M55 could be the result of the selective loss
of main sequence stars, especially from the outer envelope of the
cluster, caused by the strong tidal shocks suffered by M55 during its
many passages through the Galactic disk and near the Galactic bulge 
\cite[for a general discussion of the problem]{Piotto93}.
A flat MF for M55 agrees well with the results of \cite{CPS93} who have
found that the clusters with a small $R_{GC}$ and/or $Z_{GC}$ show a MF
significantly flatter than the cluster in the outer Galactic halo or
farther from the Galactic plane.
Indeed, M55 is near to the Galactic bulge, $R_{GC}=4.7$~kpc 
($R_\odot=8.0$~kpc), and to 
the Galactic disk $Z_{GC}=-2.0$~kpc. 
Figure~\ref{m55xrz} shows that taking into account observing errors, M55
fairly fits into the relation given by \cite{Manu97}, which is a
refined version of the one found by \cite{DPC93}.
A different conclusion has been reached by \cite{Mand96} using 
their uncorrected (for mass segregation) value for the MF of M55.
As noted by the referee, M55 lies further from the average relation
defined by the other clusters: of those with a similar abscissa
($0.0\pm0.2$), M55 is the one with the lowest value of $x$.
It is not possible to identify the main source of this apparent enhanced
mass-loss of M55 compared to the other clusters; a possible cause can be
a orbit of the cluster that deeply penetrate into the bulge of the
Galaxy.
This cannot be confirmed until is performed a reliable measure of the 
proper motion of M55.

\section{Radial density profile from star counts.\label{M55rprof}}
The CMD allows a unique way to obtain a reliable measure of the 
radial density profiles of GCs.
In fact, the CMD allows us to sort out the stars belonging to the 
cluster, limiting the problems generated by the presence of the field 
stars.
This also permits to extract radial profiles for distinct stellar 
masses.

We have first created a profile as in \cite{King68}, in order to compare
our results with the existing data in the literature.
The comparison has been done with the radial density profile of M55
published by \cite{Pryor91} which includes the visual star counts
of King et al..
We could not compare our data with \cite{Irwin84} because they
have not published their observations in tabular form.

\begin{figure*}[t]
\resizebox{12cm}{!}{\includegraphics{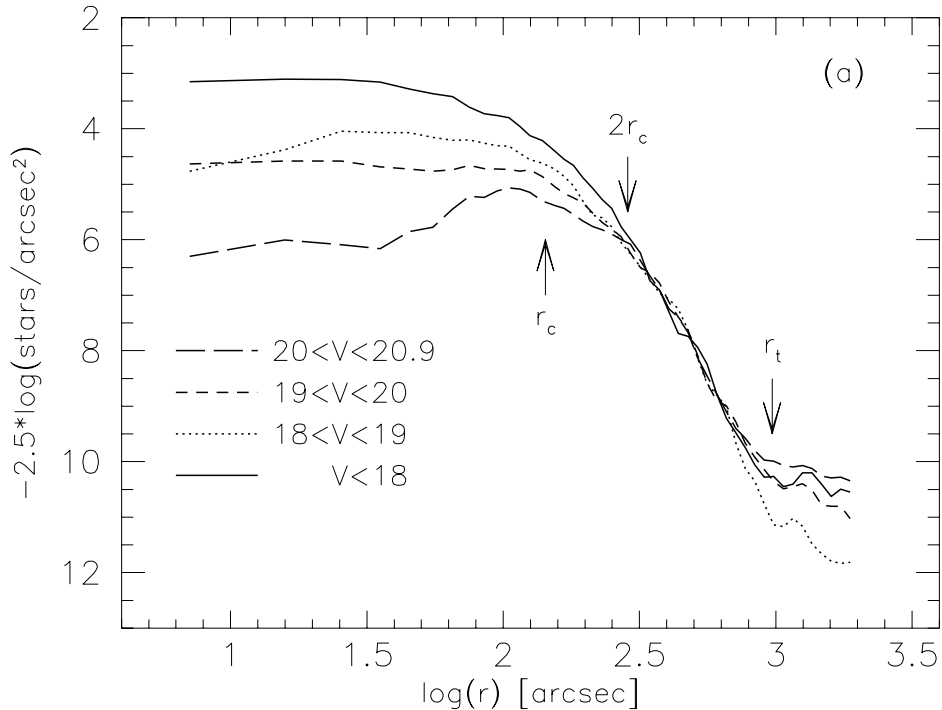}}
\hfill
\parbox[b]{55mm}{
\caption{Radial density profile of M55 for different magnitude intervals.
All the profiles has been smoothed in the external parts.
In the graph the positions of $r_c$, 2$r_c$ and $r_t$ are shown.
The profiles have been normalized in the radial range 
$2.6<\log(r)<2.9$.}}
\label{profconf}
\end{figure*}

\subsection{Density profile for stars above the TO}
Figure~\ref{profile} shows the radial density profile for the TO plus 
SGB stars extracted from the CMD of M55 (from 1 magnitude 
below the TO to the brightest limit of our photometry).
We have selected the stars within $2.5\sigma$ from the fiducial line 
of the CMD plus the contribution coming from the BS and HB stars;
star counts has been limited at the magnitude $V\le18.5$.
This relatively bright limit corresponds approximately to
the limit of the visual star counts by \cite{King68} 
on the plate ED-2134 (in order to make the comparison easier we 
used the same radial bins of King).
Our counts have been transformed to surface brightness and adjusted in 
zero point to fit the \cite{Pryor91} profile of M55.

The agreement with the data presented by Pryor is good everywhere but 
in the outer parts where our CCD star counts are clearly above those 
of \cite{King68}.
This difference is probably due to our better estimate of the
background star contamination.
In the plot we have shown also the raw star counts (crosses)
prior to the background star subtraction: it can be clearly seen that 
our star counts go well beyond the tidal radius, $r_t=977''$, published by
\cite{Trager95}.
This allow us to estimate in a better way than in the past the stellar
background contribution.
The background star counts show a small radial gradient: we will discuss 
this point in greater detail in the next Section.
Here, the minimum value has been taken as an estimate of the background
level.

We point out that the differences present in the central zones of the
cluster could be in part due to some residual incompleteness of our star
counts, to the absence in the starcounts of the brightest saturated
stars, and to the difficulties in finding the center of the cluster. 
We searched for the center using a variant of the mirror autocorrelation
technique developed by \cite{DJ88}.
In the case of M55 we encountered some problems due to a surface density
which is almost constant inside a radius of $\simeq100\arcsec$.

In order to evaluate the structural parameters of M55, we have fitted the
profile Figure~\ref{profile} with a multi-mass isotropic \cite{King66}
model as described in the previous Section.
In the following table we show the parameters of the best fitting model 
and we compare them with the results of \cite{Trager95}, 
\cite{Pryor91}, and \cite{Irwin84}:
\bigskip
\begin{center}
\begin{tabular}{lccc}
\hline\hline\noalign{\smallskip}
Author &  $c$  &  $r_c$  &  $r_t$ \\ 
\noalign{\smallskip}\hline\noalign{\smallskip}
This paper        & 0.83  & $143''$ & $970''$\\ 
Trager \etal\     & 0.76  & $170''$ & $977''$\\ 
Pryor  \etal\     & 0.80  & $140''$ & $876''$\\ 
Irwin and Trimble & $\sim1.0$ & $\sim120''$ & $\sim1200''$ \\
\noalign{\smallskip}\hline\hline
\end{tabular}
\end{center}
\bigskip
The concentration parameter of M55 is one of the smallest known for a
globular cluster.
Such a small concentration implies strong dynamical evolution and
indicates that the cluster is probably in a state of high disgregation
\cite[]{Aguilar88,Gnedin97}.

Our value of the tidal radius is well in agreement with that of 
\cite{Trager95} who used a similar method to fit the data.
\cite{Pryor91} give a value of $r_t$ 10\% smaller than ours.
We note that Pryor and Trager used the same observational data set.
The difference with \cite{Irwin84} is probably due to the fact
that the authors have not fitted their data directly but made only a
comparison with a plot of King models.

\begin{figure*}[t]
\resizebox{12cm}{!}{\includegraphics{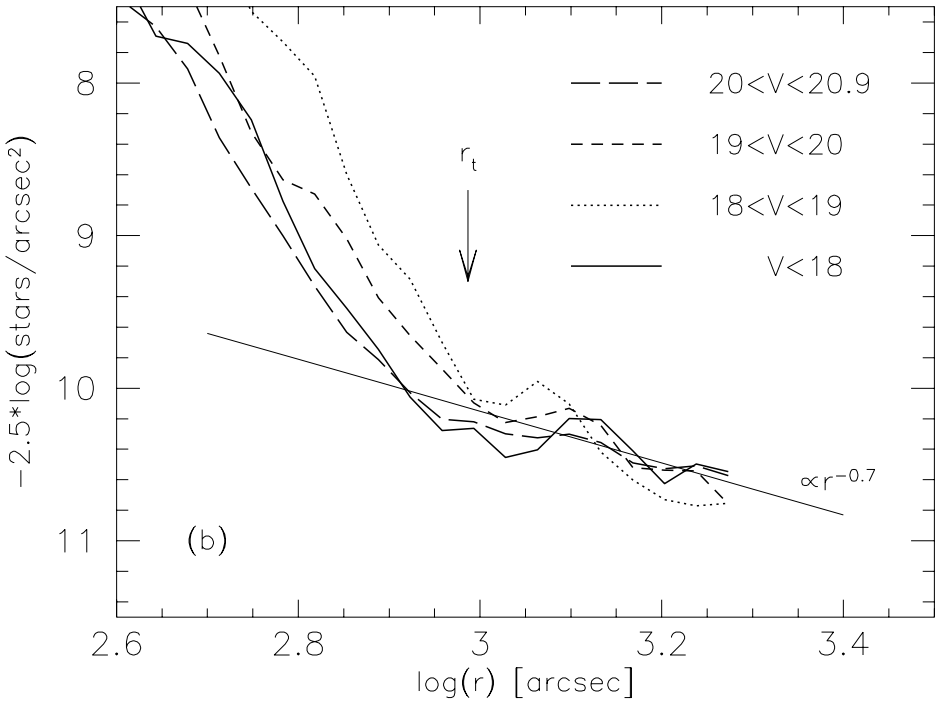}}
\hfill
\parbox[b]{55mm}{
\caption{Radial density profile of M55 for different magnitude intervals.
An arrow marks the position of $r_t$.
The profiles are normalized in the radial range $1.0r_t<r<2.0r_t$ to better 
compare the extra-tidal profiles.
We show also the power law interpolation of the profiles.}}
\label{extrat}
\end{figure*}

\subsection{The density profile for different stellar masses.}
Having verified the compatibility of our density profile with previously
published ones, we have extracted surface density profiles for different 
magnitude ranges corresponding to different stellar masses.
The adopted magnitude intervals have been chosen to have a significant 
number of stars in each bin.
We used logarithmic radial binning that allows a better sampling of
the stars in the outer part of the cluster.
In order to lower the noise in the outer part of the profile, we have
smoothed the profiles with a median static filter of fixed width
of 3 points.
We verified that the filtering procedure did not introduce spurious
radial gradients in the density profiles.
The mean masses in each magnitude bin adopted for the profiles, as 
obtained from the isochrone by \cite{VDB85} (\cf\ also
Section~\ref{Sm55mf}), are:

\begin{center}
\begin{tabular}{cc}
\hline\hline\noalign{\smallskip}
   $V$     & $<m>$ \\
\noalign{\smallskip}\hline\noalign{\smallskip}
$<18$    & 0.79 \\
$18-19$  & 0.77 \\
$19-20$  & 0.71 \\
$20-20.9$& 0.63 \\
\noalign{\smallskip}\hline\hline
\end{tabular}
\end{center}
The relative profiles, without subtraction of the back\-gr\-ou\-nd stars,
are shown in Figure~\ref{profconf} and \ref{extrat}. 
The arrows in both figures indicate $r_c$, $2~r_c$ and $r_t$.

The profiles plotted in Figure~\ref{profconf} are clearly different 
from each other: this is as expected from the mass segregation effects.
To better compare the profiles, in Figure~\ref{profconf} they have been 
normalized in the radial interval $2.6<\log(r/arcsec)<2.9$ (where the
profiles have a similar gradient) to the profile of the TO stars.
This operation is possible because in this radial range the effects of
mass segregation are small (\cf\ Figure~\ref{M55slx}); they are
more evident within one core radius.
The density profiles are consistent with the mass segregation effects 
that we have already seen in the mass function of the cluster.

\begin{figure}[t]
\resizebox{\hsize}{!}{\includegraphics{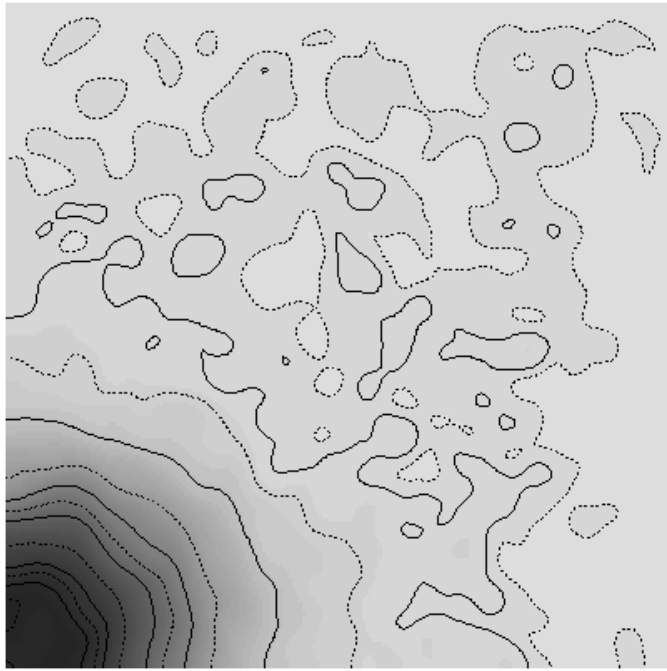}}
\caption{
Surface density map of M55 with contour levels.}
\label{gray}
\end{figure}

The more interesting aspect of the profiles in Figure~\ref{profconf} 
is the clear presence of a stellar radial gradient in the star counts 
of the background field stars.
In Figure~\ref{extrat}, we show the radial profiles of the extra cluster 
stars after normalization of the profiles outside $\log(r/arcsec)=3.0$.
The 4 profiles are not exactly coincident outside $r_t$.
Let us discuss various possible explanations for this observation:

\begin{itemize}
\item \emph{Errors in the completeness correction or errors in the star
counts.}
We repeated the extensive tests on the data made to assess the validity 
of the mass segregation seen in the LFs.
We checked that the variation in the completeness limit of the various 
EMMI fields does not introduce spurious trends.
In a different test, we divided the cluster in two slices along a line at 
$45^\circ$ from the center of the cluster up to the field~19 (\cf\ 
Figure~\ref{M55frames}), and built the radial profiles for each of the 4
magnitude bins: in all cases there were no significant differences.
The radial profile of the stars in the magnitude range $18\div19$
($M_V=4.1\div5.1$ in Figure~\ref{M55fdl}) has the lowest contamination
of background stars, as shown by its LF in Figure~\ref{M55fdl}.

\item \emph{A non-uniform distribution of the field stars around M55.}
It is possible that the field stars around M55 are distributed in a 
non-uniform way.
In the work by \cite[]{Grillmair95} it clearly appears that the field 
stars of some GCs present a non-uniform distribution around the clusters.
The gradients are significant and the authors used bidimensional
interpolation to the surface density of the field stars to subtract 
their contribution to the star counts of the clusters.
In the present case, field star gradients could be a real possibility,
but we cannot test it because we do not have $360^\circ$ coverage of the
cluster: our coverage of M55 is only a little more than a quadrant.
The Galactic position of M55 ($l\simeq-23^\circ$, $b\simeq9^\circ$) can 
give some possibility to this option.
At this angular distance from the Galactic center the bulge and halo
stars probably have a detectable radial gradient.
However it remains difficult to explain the existence of the gradient
also for the stars in the magnitude range $18\div19$: for them (\cf\
section~\ref{M55lf}), as stated before, we have the lowest
contamination from the field stars.

We have created a surface density map of the starcounts of M55. 
The map was constructed using all the stars of the $2.5\sigma$-selected
sample of our photometry (excluding fields 25 and 35),
counting stars in square areas of approximately $9''\times9''$ and then
smoothing the resulting map with a gaussian filter.
The starcounts are not corrected for crowding but we stopped at 
$V=20.5$.
The map is presented in Figure~\ref{gray}.
The map has the same orientation as Figure~\ref{M55frames}. 
We have also overplotted contour levels to help in reading the map.
Figure~\ref{gray} clearly shows that well outside the tidal radius
of M55 (located approximately at the center of the map) there is a
visible gradient in the star counts.

\item \emph{A gradient generated by the presence of the dwarf 
sphe\-roi\-dal in Sagittarius} \cite[]{Ibata95}.
Between the Galaxy center and M55 there is the dwarf spheroidal galaxy
called Sagittarius \cite[]{Ibata95}.
Sa\-git\-tarius is interacting str\-on\-gly with the Galaxy and probably 
is in the last phases of a tidal destruction by the Galactic bulge.
The distance between the supposed tidal limit of this galaxy (using the 
contour map of Ibata \etal 1995\nocite{Ibata95}) and M55 is $\sim5^\circ$.
In the recent work by \cite{Mand96} the giant sequence of the 
Sagittarius appears clearly overlapped with the sequence of M55. 
This happens only in the magnitude range $V\simeq20.0\div21.0$ 
where our star counts end.
\cite{Fahlman96} showed that the SGB sequence of the Sagittarius
crosses the main sequence of M55 at $V\simeq20.5\div20.7$, and at a
corresponding color of $(V-I)\simeq1.1\div1.2$.
Similar results were found by \cite{Mateo96}.
This is due to the different distances of these two systems from us: 
$\sim4.5$~kpc for M55 and $\sim24$~kpc for Sagittarius.
This implies that out star counts can be influenced by the stars of the 
dwarf spheroidal only in our last magnitude bin, $20\div20.9$.
Our selection of stars along the CMD of M55 limits the Sagittarius 
stars to those effectively crossing the main sequence.
In conclusion, if effectively the Sagittarius stars are present as background 
stars we should see them only in one of the 4 profiles, but the 
coincidence of the 4 profiles excludes this ipothesis.

\item \emph{A halo of stars escaping from the clusters.}
This possibility is more suggestive.
The stellar gradient could be a possible extra-tidal extension of the
cluster, similar to what \cite{Grillmair95} found in their sample of
12 clusters.
The tidal extension could be caused by the tidal-shocks to which the 
cluster has been exposed during its perigalactic passages,
through the Galactic disk.
Another possibility is the creation of the stellar halo by stellar 
dynamical evaporation from the inner part of the cluster.
Such mechanisms work independently of stellar mass
\cite[]{Aguilar88} and so the stellar halo should have a
similar gradient for all the stellar masses as in the present case.
Such halos are very similar to the theoretical results obtained by 
\cite{OhLin92} and \cite{Grillmair95}, who have obtained tidal 
tails for globular clusters N-bodies simulations.

\end{itemize}

We believe that the probable explanation for the phenomenon shown in 
Figures~\ref{profconf} and \ref{extrat} is in the presence of an
extra-tidal stellar halo or tidal tail.
Doubt resides in the unknown gradient of the background field stars.
To resolve this we need to map the whole cluster and a large
area surrounding the cluster.
This would also allow us to find the exact level of field stars.
Our star counts stop at 33\arcmin ($\simeq2\times r_t$), 
from the center of M55 while the tidal tails of \cite{Grillmair95}
stop at $\simeq 2.5 \div 4\, r_t$.
Consequently, we cannot correctly subtract the contribution
of the field stars from our star counts.
We can give only an estimate of the exponent of the power law, 
$f\propto r^{-\alpha}$, fitting the profiles at $ r > 1.2\,r_t$.
Without subtracting any background counts $\alpha\sim0.7\pm0.3$, while 
subtracting different levels of background stars the slope varies in the 
interval $0.7<\alpha<1.7$: the highest value comes out after subtracting 
the outermost value of the density profiles.
When it will be available a better estimate of the background/foreground
level of the sky it will be possible to assign a value to the slope of
the gradient of stars: actually our range, $\alpha=0.7 \div 1.7$, is in
accordance with those found theoretically by \cite{OhLin92} and
observationally by \cite{Grillmair95}.

\begin{acknowledgements}
We are grateful to C.J.~Grillmair, the referee, for his careful reading 
of the manuscript and his suggestions for improving the paper.
The authors warmly thank Tad~Pryor for making available it's code for the 
generation of multi-mass Ki\-ng-Michie models.
We thanks I.~Saviane for his help in constructing the surface density
map of M55.
Finally, we warmly thanks Nicola Caon for doing the observations 
included in this work.
\end{acknowledgements}

%\bibliographystyle{astron}
%\bibliography{mnemonic,biball}

\begin{thebibliography}{}

\bibitem[\protect\astroncite{Aguilar et~al.}{1988}]{Aguilar88}
Aguilar, L., Hut, P., \& Ostriker, J., 1988,
\newblock ApJ,  355, 720

\bibitem[\protect\astroncite{Alcaino et~al.}{1992}]{Alcaino92}
Alcaino, G., Liller, W., Alvarado, F., \& Wenderoth, E., 1992,
\newblock AJ,  104, 190

\bibitem[\protect\astroncite{Bailyn}{1995}]{Bailyn95}
Bailyn, C., 1995, 
\newblock ARAA, 33, 133

\bibitem[\protect\astroncite{Capaccioli et~al.}{1993}]{CPS93}
Capaccioli, M., Piotto, G., \& Stiavelli, M., 1993,
\newblock MNRAS,  261, 819

\bibitem[\protect\astroncite{Djorgovski}{1988}]{DJ88}
Djorgovski, S., 1988,
\newblock in The Harlow-Shapley Symp. on Globular Cluster Systems in Galaxies;
  eds. J.E. Grindlay and A.G. Davis Philip, Proceedings of the IAU Symposium
  126, (Dordrecht: Reidel), p. 333

\bibitem[\protect\astroncite{Djorgovski}{1993}]{DJ93}
Djorgovski, S., 1993,
\newblock in ASP Conf. Ser., Vol. 50, Structure and Dynamics of Globular
Clusters, ed. S.Djorgovski \& G.Meylan (San Francisco: ASP), p. 373

\bibitem[\protect\astroncite{Djorgovski \& Piotto}{1993}]{DP93}
Djorgovski, S., Piotto, G., 1993,
\newblock PASPC, 50, 373

\bibitem[\protect\astroncite{Djorgovski et~al.}{1993}]{DPC93}
Djorgovski, S., Piotto, G., \& Capaccioli, M., 1993,
\newblock AJ,  105, 2148

\bibitem[\protect\astroncite{Fahlman et~al.}{1996}]{Fahlman96}
Fahlman, G. G., Mandushev, G., Richer, H. B., Thompson, I. B.,
Sivaramakrishnan, A., 1996,
\newblock ApJ, 459, L65

\bibitem[\protect\astroncite{Ferraro et~al.}{1997}]{Ferraro97}
Ferraro, F.~R., \etal, 1997, 
\newblock A\&A, in press

\bibitem[\protect\astroncite{Gnedin \& Ostriker}{1997}]{Gnedin97}
Gnedin, O. \& Ostriker, J., 1997,
\newblock ApJ, 474, 223

\bibitem[\protect\astroncite{Grillmair et~al.}{1995}]{Grillmair95}
Grillmair, C.~J., Freeman, K.~C., Irwin, M., \& Qinn, P.~J., 1995,
\newblock AJ,  109, 2553

\bibitem[\protect\astroncite{Gunn \& Griffin}{1979}]{GG79}
Gunn, J. E., \& Griffin, R. F., 1979,
\newblock AJ, 84, 752

\bibitem[\protect\astroncite{Ibata et~al.}{1995}]{Ibata95}
Ibata, R.~A., Gilmore, G., \& Irwin, M.~J., 1995,
\newblock MNRAS,  277, 781


\bibitem[\protect\astroncite{Irwin \& Trimble}{1984}]{Irwin84}
Irwin, M.~J. \& Trimble, V., 1984,
\newblock AJ,  89, 83

\bibitem[\protect\astroncite{King}{1966}]{King66}
{King}, I.~R., 1966,
\newblock AJ,  71, 64

\bibitem[\protect\astroncite{King et~al.}{1968}]{King68}
{King}, I.~R., Hedeman, E., Hodge, S., \& White, R., 1968,
\newblock AJ,  73, 456

\bibitem[\protect\astroncite{Lee}{1977}]{Lee77}
Lee, S. W., 1977
\newblock A\&AS, 29, 1

\bibitem[\protect\astroncite{Lehman \& Sholz}{1997}]{LS97}
Lehman, I., \& Sholz, R. D., 1997
\newblock A\&A, in pubblication

\bibitem[\protect\astroncite{Mandushev et~al.}{1996}]{Mand96}
Mandushev, G. I. and Fahlman, G. G. and Richer, H. B., 1996
\newblock AJ, 112, 1536

\bibitem[\protect\astroncite{Mateo}{1995}]{Mateo95}
Mateo, M., 1995,
\newblock in ASP Conf. Ser., Vol. , Binaries in Clusters, 
ed. E. Milone, (San Francisco: ASP), p.  

\bibitem[\protect\astroncite{Mateo et~al..}{1996}]{Mateo96}
Mateo, M., Mirabal, N., Udalski, A., Szymanski, M., Kaluzny, J., Kubiak, 
M., Krezminski, W., \& Stanek, K.Z., 1996,
\newblock ApJ, 458, L13

\bibitem[\protect\astroncite{Oh \& Lin}{1992}]{OhLin92}
Oh, K.~S. \& Lin, D. N.~C., 1992,
\newblock ApJ,  386, 519

\bibitem[\protect\astroncite{Piotto et~al.}{1990a}]{Piotto90a}
Piotto, G., King, I., Capaccioli, M., Ortolani, S., Djorgovski, S., 1990,
\newblock AJ, 350, 662

\bibitem[\protect\astroncite{Piotto et~al.}{1990b}]{Piotto90b}
Piotto, G., King, I.~R., \& Djorgovski, S., 1990,
\newblock AJ, 96, 1918

\bibitem[\protect\astroncite{Piotto}{1993}]{Piotto93}
Piotto, G., 1993,
\newblock in ASP Conf. Ser., Vol. 50, Structure and Dynamics of Globular
  Clusters, ed. S.Djorgovski \& G.Meylan (San Francisco: ASP), p. 233

\bibitem[\protect\astroncite{Piotto et~al.}{1997}]{Piotto97}
Piotto, G., \etal, 1997,
\newblock preprint

\bibitem[\protect\astroncite{Pryor et~al.}{1991}]{Pryor91}
{Pryor}, C., {McClure}, R.~D., {Fletcher}, J.~M., \& {Hesser}, J.~E., 1991,
\newblock AJ,  102, 1026

\bibitem[\protect\astroncite{Pryor et~al.}{1986}]{Pryor86}
Pryor, C., Smith, G.~H., McClure, R.~D., 1986,
\newblock AJ, 92, 1358

\bibitem[\protect\astroncite{Renzini \& Fusi~Pecci}{1988}]{RFP88}
Renzini, A. \& Fusi~Pecci, F., 1988,
\newblock ARA\&A,  26, 199

\bibitem[\protect\astroncite{Richer \& Fahlman}{1989}]{Richer89}
Richer, H.~B. \& Fahlman, G.~G., 1989,
\newblock ApJ,  339, 178

\bibitem[\protect\astroncite{Rosenberg et~al.}{1996}]{Alf96}
Rosenberg, A., Saviane, I., Piotto, G., Zaggia, S. R., 
\& Aparicio, A., 1996,
\newblock in Dynamical Evolution of Star Clusters -- Confrontation of 
Theory and Observations, eds. P. Hut \& J. Makino, IAU Symp. 174, p. 341

\bibitem[\protect\astroncite{Rosenberg et~al.}{1997}]{Alf97}
Rosenberg, A., Saviane, I., Piotto, G., Aparicio, A., \& Zaggia, S. R., 
1997,
\newblock AJ, in pubblication

\bibitem[\protect\astroncite{Saviane et~al.}{1995}]{Ivo95}
Saviane, I., Piotto, G., Fagotto, F., Zaggia, S. R., Capaccioli, M., 
\& Aparicio, A., 1995
\newblock in The Formation of the Milky Way, eds. E.J. Alfaro \& 
A.J.Delgado, Cambridge University Press, p. 301 

\bibitem[\protect\astroncite{Saviane et~al.}{1997}]{Ivo97}
Saviane, I., Piotto, G., Fagotto, F., Zaggia, S. R., Capaccioli, M., \& 
Aparicio, A., 1997
\newblock A\&A, submitted

\bibitem[\protect\astroncite{Shade et~al.}{1988}]{Shade88}
Shade, D., VandenBerg, D.~A., \& Hartwick, F., 1988,
\newblock AJ,  96, 1632

\bibitem[\protect\astroncite{Stetson}{1987}]{Stetson87}
Stetson, P., 1987,
\newblock PASP,  99, 191

\bibitem[\protect\astroncite{Trager et~al.}{1995}]{Trager95}
Trager, S.~C., King, I.~R., \& Djorgovski, S., 1995,
\newblock AJ,  109, 218

\bibitem[\protect\astroncite{VandenBerg \& Bell}{1985}]{VDB85}
VandenBerg, D.~A. \& Bell, R.~A., 1985,
\newblock ApJS,  58, 561

\bibitem[\protect\astroncite{Veronesi et~al.}{1996}]{Carla96}
Veronesi, C., Zaggia, S.R., Piotto, G., Ferraro, F.R., \& Bellazzini, 
M., 1996,
\newblock in Formation of the Galactic Halo... Inside and Out, A.S.P.
Conference Series, vol. 92, eds. H. Morrison \& A. Sarajedini, p.301

\bibitem[\protect\astroncite{Weideman}{1990}]{W90}
Weideman, V., 1990,
\newblock ARAA, 28, 103

\bibitem[\protect\astroncite{Zaggia et~al.}{1995}]{Zaggia95}
Zaggia, S.~R., Piotto, G., \& Capaccioli, M., 1995,
\newblock Mem. Soc. Astron. Ital.,  441, 667

\bibitem[\protect\astroncite{Zinn}{1980}]{Zinn80}
Zinn, R., 1980,
\newblock ApJS,  42, 19

\bibitem[\protect\astroncite{Zoccali et~al.}{1997}]{Manu97}
Zoccali, M., Piotto, G., Zaggia, S.~R., \& Capaccioli, M., 1997,
\newblock MNRAS, submitted

\end{thebibliography}

\newpage

\appendix

\begin{table*}[t]
\caption{
For each field of M55 we list the total number of detected stars in both
the $V$ and $I$ frame, the mean airmass of the field, the right ascension
and the declination of the field center, the FWHM of the $V$ and $I$ point
spread functions of the images, and the $V$ limit magnitude of the
observed fields. For each $V$ image the exposure time was of 40 seconds,
while for the $I$ image it was of 30 seconds. \label{M55tab2}}
\begin{tabular}{crcccccc}
\hline\hline\noalign{\smallskip}
Field&Stars&Airmass&    RA   &    DEC    &\multicolumn{2}{c}{FWHM} & V(50\%)\\
     &     &       &\multicolumn{2}{c}{[deg]}&    $V$   &  $I$    &   \\
\noalign{\smallskip}\hline\noalign{\smallskip}
01 & 13442 & 1.010 & 294.926 & $-30.948$ &  0.9   & 0.9   &  21.0/22.1 \\
02 &  3176 & 1.007 & 295.062 & $-30.948$ &  0.9   & 0.9   &  22.2 \\
03 &   872 & 1.004 & 295.198 & $-30.948$ &  1.1   & 1.1   &  22.1 \\
04 &   495 & 1.005 & 295.333 & $-30.947$ &  1.3   & 1.1   &  21.2 \\
06 &  2273 & 1.016 & 294.926 & $-30.831$ &  1.3   & 1.2   &  21.5 \\
07 &   800 & 1.013 & 295.061 & $-30.831$ &  1.5   & 1.5   &  21.2 \\
08 &   543 & 1.009 & 295.197 & $-30.831$ &  1.4   & 1.4   &  21.2 \\
09 &   482 & 1.007 & 295.333 & $-30.831$ &  1.6   & 1.6   &  21.0 \\
11 &   635 & 1.022 & 294.926 & $-30.714$ &  1.4   & 1.4   &  21.3 \\
12 &   560 & 1.028 & 295.061 & $-30.714$ &  1.4   & 1.4   &  21.2 \\
13 &   531 & 1.034 & 295.197 & $-30.714$ &  1.4   & 1.5   &  21.0 \\
14 &   506 & 1.041 & 295.333 & $-30.714$ &  1.3   & 1.5   &  21.0 \\
16 &   491 & 1.227 & 294.983 & $-30.598$ &  1.3   & 1.4   &  20.9 \\
17 &   503 & 1.150 & 295.061 & $-30.598$ &  1.6   & 1.3   &  20.9 \\
18 &   537 & 1.055 & 295.197 & $-30.598$ &  1.5   & 1.6   &  21.1 \\
19 &   556 & 1.048 & 295.332 & $-30.598$ &  1.3   & 1.5   &  21.2 \\
25 &   666 & 1.001 & 295.197 & $-31.063$ &  1.2   & 1.3   &  21.6 \\
35 &   757 & 1.001 & 295.333 & $-31.063$ &  1.2   & 1.1   &  21.9 \\
\noalign{\smallskip}\hline\hline
\end{tabular}
\end{table*}

\section{Image reduction and analysis}

The images were reduced using the standard algorithms 
of bias subtraction, flat fielding and trimming of the overscan, without 
encountering particular problems.
The stellar photometry was done using 
{\tt DAOPHOT II} and {\tt ALLSTAR} \cite[]{Stetson87}.
The second version of {\tt DAOPHOT} was particularly useful for the 
image analysis since we were forced to use a variable 
point spread function (PSF) through the images.
In fact, the stellar images of the EMMI \emph{Red Arm} together with the
F/2.5 field camera presented coma aberration at the edges of the field:
the resulting PSF was radially elongated.
To better interpolate the PSF we also used an analytic function with 
5 free parameters (\ie\ the {\tt Penny} function of {\tt DAOPHOT II}).

In order to obtain a single CMD for all the stars found in the 18 
fields we first obtained the CMD of each field matching the $V$ and I
photometry.
Then, we combined all the CMDs using the relative zero points determined
from the overlapping regions of adjacent fields.
All the CMDs were connected to the main CMD, one at a time, following a
sequence aimed at maximizing the number of common stars usable 
for the zero point calculation.
The central field CMD was used as the starting point of the combination.
For the outer fields we used a minimum of 20 common stars while for 
the inner fields we had at least 300 stars.
The mean error of the zero points was $\simeq0.05$ magnitudes, compatible
with the errors calculated from the crowding experiments.
Since the night was not photometric, we could not directly calibrate 
our data. 
We were only able to set an absolute zero point using the unpublished
calibrated photometry of the center of M55 by Piotto (see next Section).

In order to perform the photometry of the central field of the cluster, we
divided it into 4 subimages of $\simeq600\times600$ pixels, to minimize
the effect of the strong stellar gradients present in this image.
We allowed a good overlap to be able to perform the successive combination 
of the photometry of the stars.
In this way we also avoided two problems: we had better control of the PSF 
calculation and we reduced the number of stars per image to be analyzed.
Thanks to the low central concentration of the cluster and the fairly
good seeing of the images (even if the crowding was not completely
absent), we were able to obtain complete photometry down to
$V\simeq21$.

\begin{figure}[t]
\resizebox{\hsize}{!}{\includegraphics{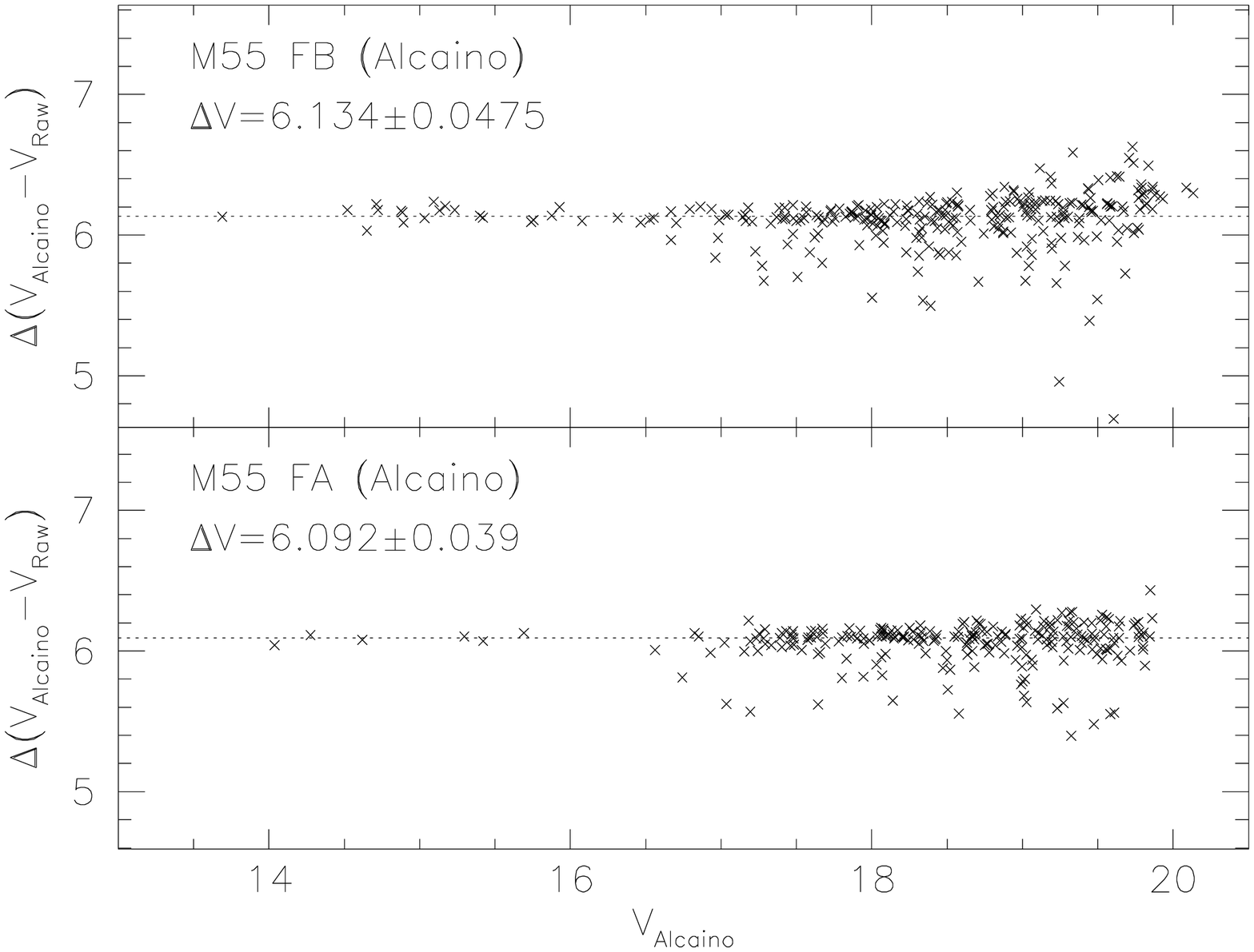}}
\resizebox{\hsize}{!}{\includegraphics{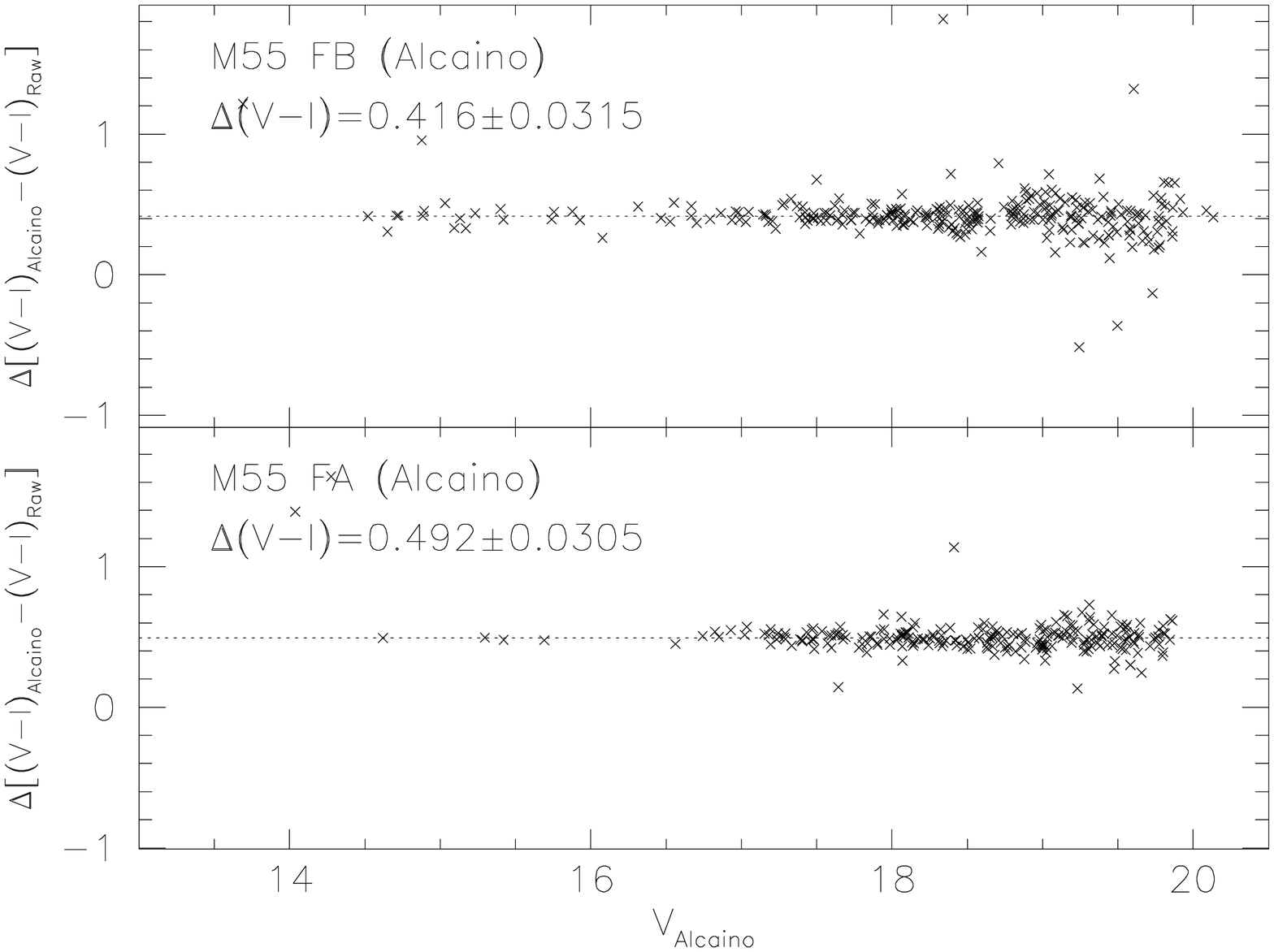}}
\caption{ \emph{top:}
Differences between our instrumental $V$ magnitude ($V_{raw}$)
and the calibrated magnitude by Alcaino \etal (1992) ($V_{Alcaino}$).
\emph{Bottom:}
Differences between our instrumental color, $(V-I)_{raw}$,
and the calibrated color by Alcaino (1992), $(V-I)_{Alcaino}$.} 
\label{M55Alcv}
\end{figure}
\begin{figure}
\resizebox{\hsize}{!}{\includegraphics{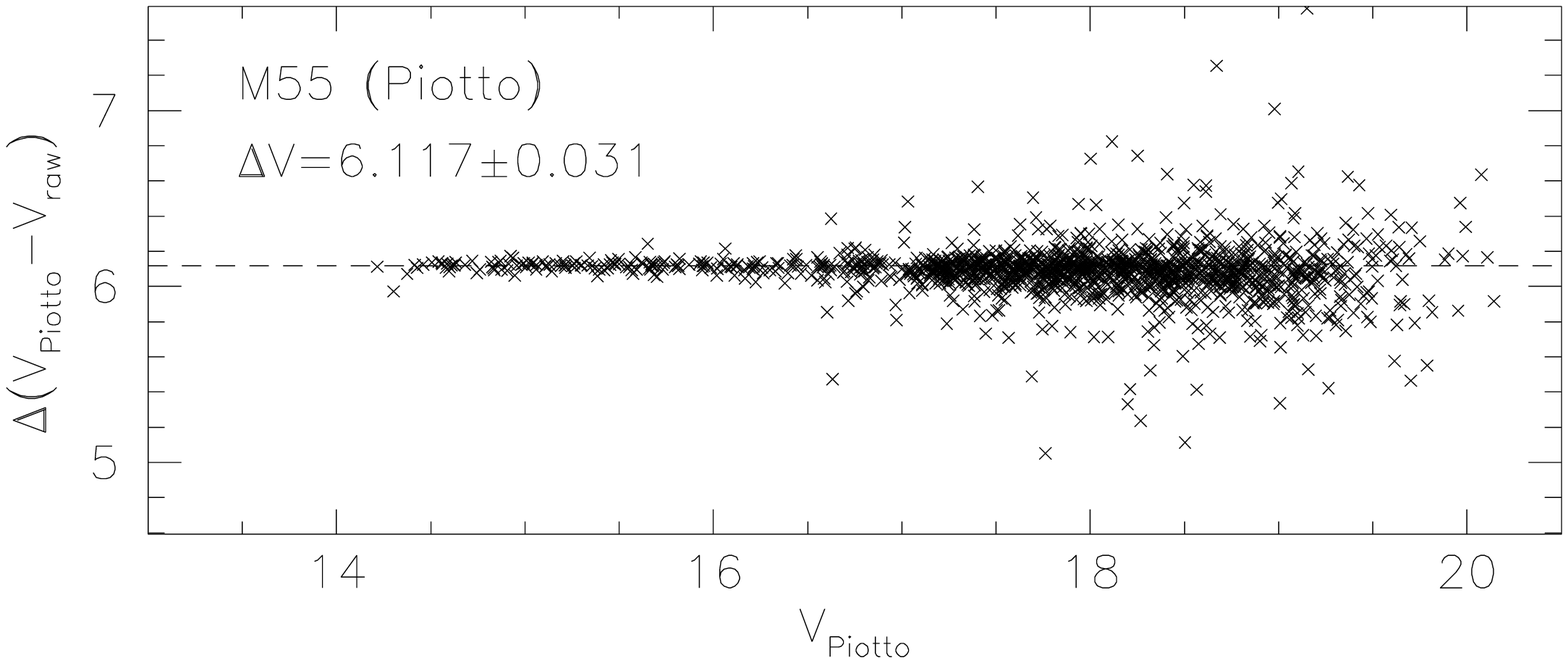}}
\caption{
Differences between our instrumental $V$ magnitude ($V_{raw}$)
and the calibrated magnitude of Piotto (1994) ($V_{Piotto}$).}
\label{M55P8}
\end{figure}

\section{Calibration of the photometry}
In principle, the analysis of the radial density profile does not require 
calibrated photometry.
But this operation is necessary if we want to analyze the stellar population 
of the cluster, together with its stellar luminosity and mass functions.
Since we could not use standards taken during the same night, we have 
performed a relative calibration using existing photometry of M55.
For the $V$ magnitude we linked our data to Piotto's (1994) un published
photometry of the central field of M55 from images taken with 
the 2.2\,m~ESO telescope.
For the (V-I) we calibrated our data against \cite{Alcaino92}  
photometry.
They published a CCD BVRI photometry for two different non-over\-lap\-ping 
fields outside the center of M55, named FA and FB, 
with dimensions of $3\farcm1\times1\farcm9$, contained in our central 
field.
Figure~\ref{M55P8} shows our $V$ zero point calculated against Piotto's (1994)
while Figure~\ref{M55Alcv} shows the $V$ zero point of the two fields 
FA and FB of \cite{Alcaino92}.
The mean zero point for the two fields of Alcaino \etal\ gives 
$\Delta V_{Alcaino}=6.11\pm0.04$ which compare well with 
$\Delta V_{Piotto}=6.12\pm0.03$.
The two are in good agreement taking into account the errors.
There are no magnitude gradients.
The LFs are coming from the photometry in the V-band.

Before the publication of the I-band photometry by \cite{Mand96},
the one by \cite{Alcaino92}  was the only photometry in the
literature.
Unfortunately, the M55 data set of \cite{Mand96} does not
overlap with any of our fields: it is centered just few arcmin south of
our field 2.
In Figure~\ref{M55Alcv}, we show the difference between our data and 
those of \cite{Alcaino92}. 
In this case, the two zero points calculated for Alcaino's 
fields differ by a significant amount.
We do not know the origin of this discrepancy, which we believe 
is internal to the data of \cite{Alcaino92}.
They could not resolve this due to the fact that fields
FA and FB do not have stars in common.
We believe that the problem is not in our data since both Alcaino's 
fields are contained in the same subimage of the central field.
Lacking other independent (V-I) calibrations, we are forced to adopt as
our color zero point the mean of the two values of FA and FB:
$\Delta(V-I)=0.45\pm0.05$.

\section{Crowding experiments.}\label{crowd}
For each field, we performed a series of Monte~Carlo simulations in order
to establish the magnitude limit and the degree of completeness of the
CMD.
The magnitude limit has been defined as the level at which the completeness 
function reach a value of $0.5$, V(50\%).
This value is reported in Table~\ref{M55tab2} for each field.

The procedure followed to generate the artificial stars for the 
crowding experiments is the standard one \cite[]{Piotto90b}.
The completeness function used to correct our data is the combination of 
the results of the experiments in both the $V$ and $I$ images for each field.
In the outer fields the stars were added at random positions in a 
magnitude range starting from $V=19$ (just 0.5 mag below the 
main sequence TO).
In the $I$ band experiments, we used the same star positions of the $V$ 
experiments, with the $I$ magnitudes set according to the corresponding
main sequence color.
For each outer field, we performed 10 experiments with 100 stars.
For the inner fields (fields number 2, 3 and 6) the experiments were 10 
with 100 stars in an interval of only 1 magnitude for 5 different 
magnitudes (a total of 50 experiments).
Moreover, in these fields the stars were added taking into account the radial 
density profile of the cluster.
For the 4 subimages of the central field, we performed independent
crowding experiments.
For each subimage, we ran 10 experiments in 0.5 mag. steps in the range
$19\div23$, with the stars radially distributed as the density profile
of the cluster.
In this way, we were able to better evaluate the level of the local
completeness of the photometry.

\begin{figure}[hb]
\resizebox{\hsize}{!}{\includegraphics{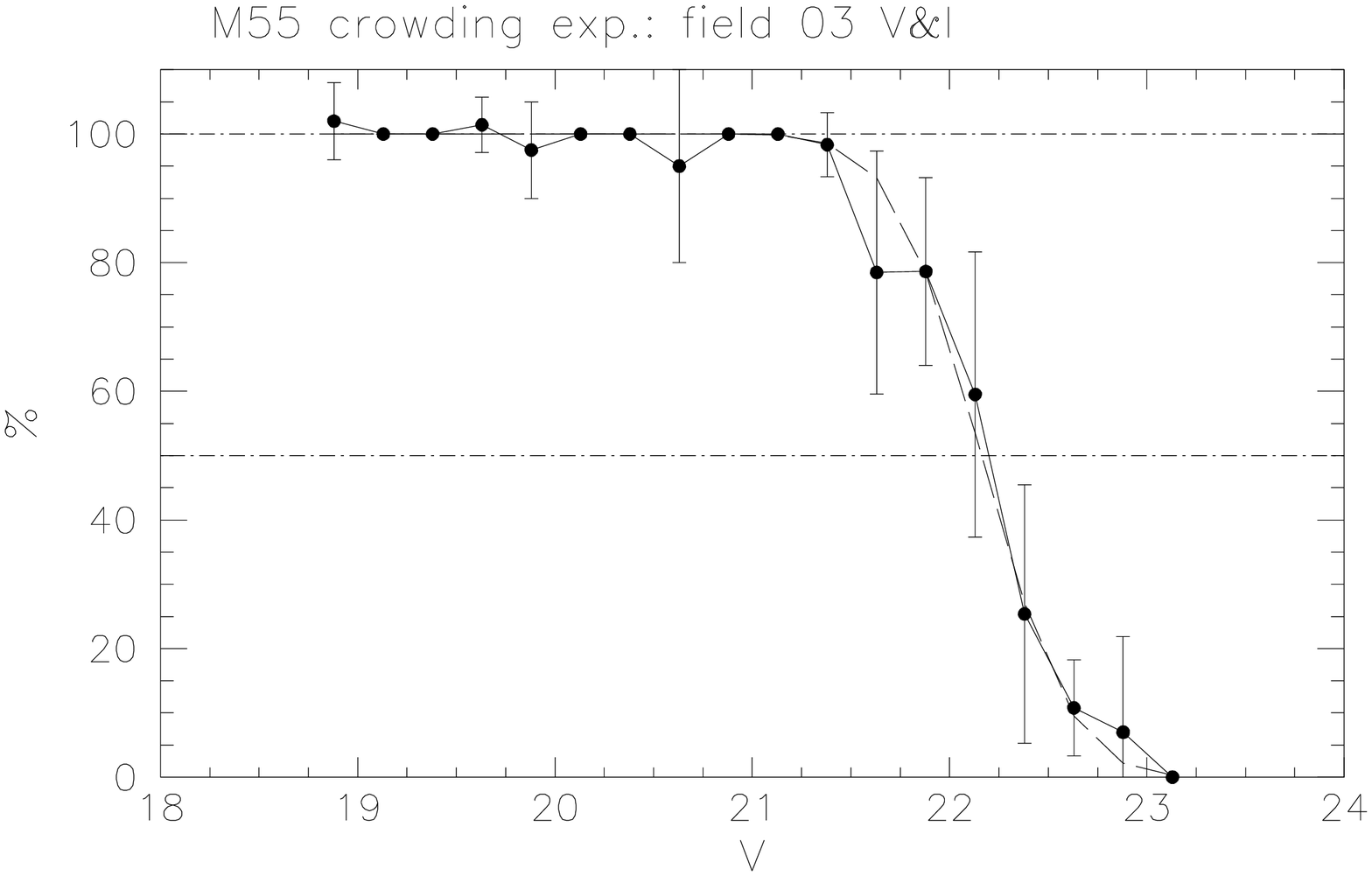}}
\bigskip
\resizebox{\hsize}{!}{\includegraphics{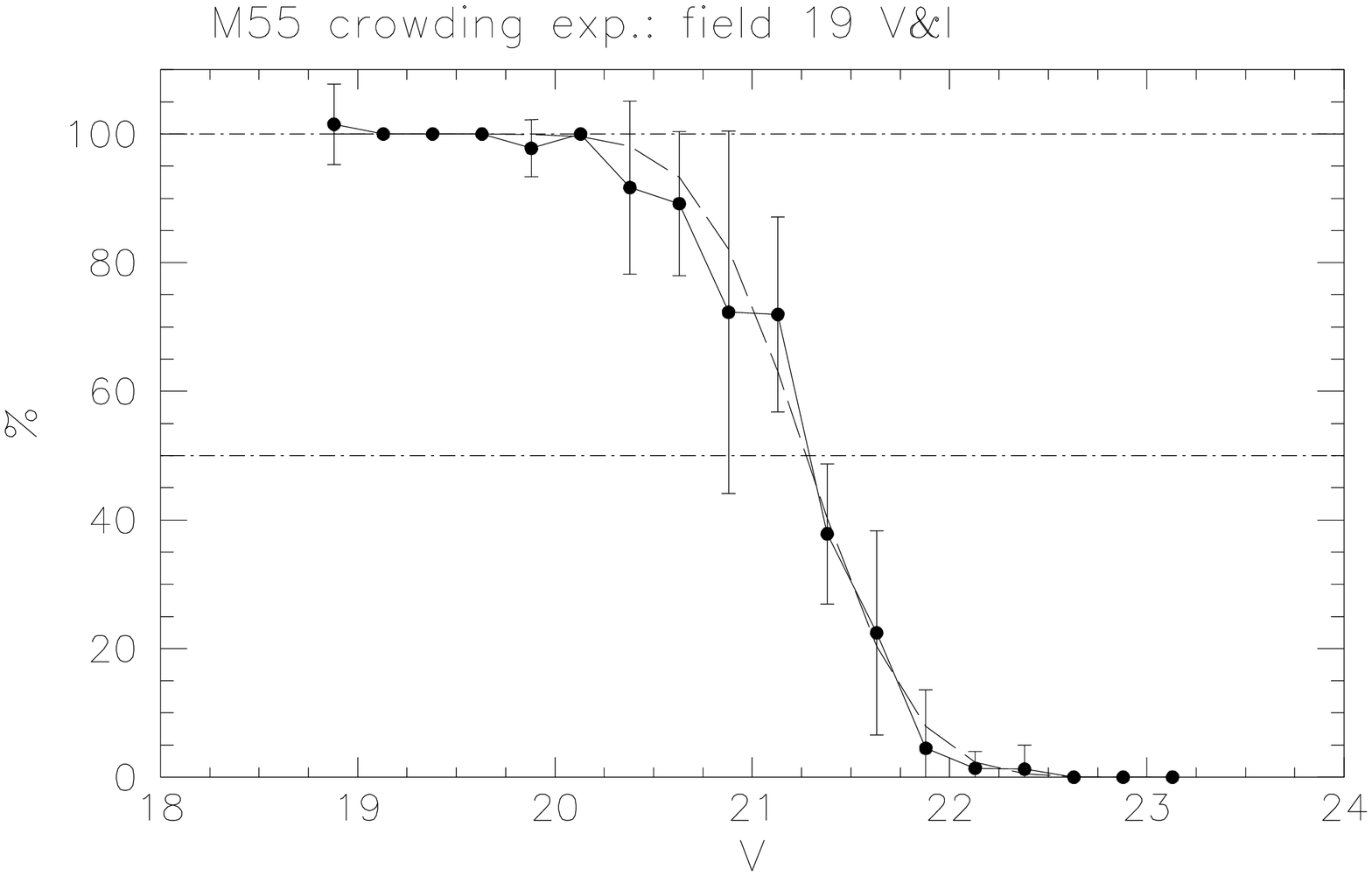}}
\caption{\emph{Top:} Completeness function for the field number 3.
\emph{Bottom:} Completeness function for the field number 19.
Both functions have been obtained combining the crowding experiments on
in $V$ and I.}
\label{M55cro}
\end{figure}

The completeness function has been calculated for each field taking into 
account the results of the two different experiments in $V$ and I.
As an example in Figure~\ref{M55cro} we show the completeness functions 
for field number 3 (top) and 19 (bottom).
The results of the experiments were fitted using the error function:
\begin{equation}
g(x;y_0,\sigma) = 1 - \int_{-\infty}^{x} e^{\frac{(y-y_0)^2}{\sigma^2}} 
{\rm d}y.
\label{cumgauss}
\end{equation}
$y_0$ is the magnitude at which the completeness level is 50\%, V(50\%); 
$\sigma$ gives the rapidity of the decrease of the incompleteness function 
and is connected with the read out noise and the crowding 
of the image.
For the star counts correction we used the interpolation with the
previous equation instead of using directly the noisy results of the
experiments (these were too few to lower the small number statistical
noise of the results).
In this way we avoid the adding of noise to the star counts.
In every case we verified that the fitting function is an acceptable 
interpolation that gives very low residuals compared to the error 
distribution function.

\end{document}